\documentclass[aps,%
final,%
notitlepage,%
oneside,%
twocolumn,
nobibnotes,%
nofootinbib,%
superscriptaddress,%
showpacs,%
centertags,
floatfix]%
{revtex4}
\usepackage[dvips]{graphicx}
\usepackage{epsfig,graphics}
\usepackage{float}
\usepackage[caption = false]{subfig}
\usepackage{hyperref}
\usepackage{amsmath,amsthm,amssymb,amscd}
\DeclareMathOperator{\sgn}{sgn}

\begin{document}

\title{Generation of monocycle squeezed light in chirped quasi-phase-matched nonlinear crystals}

\author{D.~B.~Horoshko}\email{Dmitri.Horoshko@univ-lille1.fr}
\affiliation{Univ. Lille, CNRS, UMR 8523 - PhLAM - Physique des Lasers Atomes et Mol\'ecules, F-59000 Lille, France}
\affiliation{B.~I.~Stepanov Institute of Physics, NASB, Nezavisimosti Ave.~68, Minsk 220072 Belarus}%

\author{M.~I.~Kolobov}
\affiliation{Univ. Lille, CNRS, UMR 8523 - PhLAM - Physique des Lasers Atomes et Mol\'ecules, F-59000 Lille, France}

\date{\today}

\begin{abstract}
We present a quantum theory of parametric down-conversion of light in chirped quasi-phase-matched second-order nonlinear crystals with undepleted quasi-monochromatic pump. This theory allows us to consider generation of ultrabroadband squeezed states of light and is valid for arbitrary, sufficiently slowly-varying nonlinear poling profiles. Using a first-order approximate quantum solution for the down-converted light field, we calculate the squeezing spectra and the characteristic squeezing angles. We compare the approximate solutions with the exact and numerical ones and find a very good agreement. This comparison validates our approximate solution in the regime of moderate gain, where the existing approaches are not applicable. Our results demonstrate that aperiodically poled crystals are very good candidates for generating ultrabroadband squeezed light with the squeezing bandwidth covering almost all the optical spectrum and the correlation time approaching a single optical cycle.
\end{abstract}

\maketitle

\section{Introduction}

Squeezed light is a non-classical electromagnetic field at optical frequency with the fluctuations of one quadrature component below the level of the vacuum fluctuations within certain frequency bandwidth. Squeezed light is one of the central objects of study in modern quantum optics, being, on the one hand, a macroscopic object with substantially quantum properties, and, on the other hand, a valuable resource for metrology, quantum communication and quantum information processing \cite{Kilin01}. Both the degree of squeezing and the squeezing bandwidth are important for potential applications of squeezed light. To date, successful generation has been reported of continuous-wave optical beams with 15 dB squeezing in a band of about 100 MHz \cite{Vahlbruch16} and 2 dB in a band of 1.2 GHz \cite{Ast13}. Experiments with pulsed light reach the bandwidth of several THz \cite{Spalter98,Wenger04,Agafonov10,Pinel12} and even tens of THz \cite{Iskhakov09}.

In our recent paper \cite{Horoshko13} we gave a theoretical description of a method allowing for generation of squeezed light with the squeezing bandwidth comprising the whole optical spectrum, i.e., hundreds of THz. After a proper compensation of the phase, such light would demonstrate a monocycle two-mode squeezing with the sideband-frequency quadrature components quantum-correlated at the time scale of a single optical period. The proposed method is based on parametric down-conversion (PDC) of light in an aperiodically poled nonlinear crystal with quasi-phase-matching (QPM) in a broad band of frequencies, resulting from a linear chirp of the spatial frequency of the poling. Such crystals are widely used for parametric amplification of ultrashort pulses of light \cite{Baker03,Fejer05,Fejer08a,Fejer08b,Heese10,Phillips14} and also for generation of photon pairs with a correlation time of the order of one optical cycle \cite{Harris07,Nasr08,Brida09,Sensarn10,Tanaka12}.

In the present article we develop a general quantum theory of generation of ultrabroadband squeezed light by PDC of light in aperiodically poled crystals. Our purpose is twofold. First, we provide a detailed description for the analytic solution of the case of linear chirp, presented in Ref.~\cite{Horoshko13}, where many details were omitted. Second, we present an approximate solution for the quantum field in a QPM nonlinear crystal, which is important for qualitative understanding of the underlying physical processes and for crystal design in practical applications. Our theory is valid for arbitrary nonlinear poling profiles, which should be sufficiently slowly varying, and for both low and high parametric gains. We compare the approximate analytical solution with exact and numerical ones for linear and quadratic-hyperbolic QPM poling profiles and find a very good agreement within the amplification band.

Our approach is conceptually close to the classical description of optical parametric amplification in QPM media developed in Refs.~\cite{Fejer08a,Fejer08b}. We use a similar perturbation approach for obtaining an approximate solution of the wave equation for the slowly-varying field amplitudes. We restrict our consideration to the first-order approximation; however our results can be easily generalized to the second-order solution. The main difference between our approach and that of Ref.~\cite{Fejer08a} is that our solution is for the slowly-varying Heisenberg field operators and, therefore, can be applied to arbitrary quantum states of light such as squeezed or entangled states. The solution of Ref.~\cite{Fejer08a} is for the classical slowly-varying field amplitudes and is not suitable for evaluation of the squeezing spectra and squeezing angles of the ultrabroadband squeezed light, which is the main objective of our work.

It should be understood also that the classical and the quantum theories of PDC in aperiodically poled crystals are oriented at different values of the parametric gain and put different meaning into the term ``high-gain regime''. For the classical theory of parametric amplification the gain is ``high'' if it provides a practically important increase of the signal peak power, above 10 dB, sometimes even above 60 dB \cite{Jovanovic05}. In the quantum theory of PDC the ``low-gain regime'' corresponds to spontaneous emission of the downconverted photon pairs, so-called biphotons, while the ``high-gain regime'' corresponds to stimulated emission of photons, when the mean photon number per mode well surpasses unity. The latter regime is characterized by squeezing of one field quadrature and can be observed at the values of the power gain, which are not practical for the pulse amplification. Indeed, for the power gain $G$ the variance of the squeezed quadrature is reduced $[G^{\frac12}+(G-1)^{\frac12}]^2$ times below the vacuum level. Thus, the widely available values of squeezing from 3 to 12 dB correspond to the power gain from 0.5 to 7 dB, which is of relatively little interest for the purpose of amplification of light pulses. We note in this connection, that our quantum solution gives an adequate description of the field evolution in the high (above 0.5 dB) and very high (above 10 dB) gain regimes, and in the latter case is in good agreement with the classical formulas obtained in Ref.~\cite{Fejer08a}.

The article is organized as follows. In Sec.~\ref{PDC} we derive a differential equation for the slowly-varying Heisenberg field operators of the electromagnetic field in an aperiodically poled nonlinear crystal. This equation is solved exactly for a linear poling profile in Sec.~\ref{linear} and approximately for an arbitrary sufficiently slowly varying poling profile in Sec.~\ref{approx}. An example of a crystal with more than octave-wide QPM is considered in Sec.~\ref{spectrasection}, where we compare the exact analytical solution for a linear poling profile with the approximate one. In the same section a similar comparison is presented for numerical and approximate analytical solutions for a nonlinear, quadratic-hyperbolic poling profile. Here we discuss also the limits of applicability of our analytical approximation. In Sec.~\ref{conclusion} we summarize the results and discuss their importance for the experiments with ultrabroadband squeezed light.

\section{Parametric down-conversion in an aperiodically poled nonlinear crystal \label{PDC}}

\subsection{Differential equation for the field}
We consider the process of collinear PDC in a nonlinear crystal, where after annihilation of one photon of the pump wave with the frequency $\omega_{\mathrm{p}}$ two photons are created with the same polarization and frequencies $\omega_{\mathrm{0}}+\Omega $ and $\omega_{\mathrm{0}}-\Omega $, where $\omega_{\mathrm{0}}=\omega_{\mathrm{p}}/2$. The phase mismatch for this process has the form $\Delta(\Omega )=k_{\mathrm{p}}-k(\Omega )-k(-\Omega )$, where $k_{\mathrm{p}}$ is the wave vector of the pump wave, accepted to be an undepleted monochromatic plane wave, and $k(\Omega )$ is the wave vector of the down-converted wave at the frequency $\omega_{\mathrm{0}}+\Omega $. In general there is no phase matching at degeneracy, $k_p\neq 2k_0$, where $k_0=k(0)$. Let us direct the $z$ axis along the propagation of the waves, placing the origin on the front edge of the crystal. For the description of the field we use two operators: the photon annihilation operator at the frequency $\omega_{\mathrm{0}}+\Omega $ and position $z$, which we denote $b\left(\Omega ,z\right)$, and the sideband photon annihilation operator at detuning $\Omega$, which is given by $a(\Omega,z)=b( \Omega,z)e^{i\left( k( \Omega)-k_0\right)z}$. The field operator $E^{(+)}(t,z)$ of the down-converted light is expressed (in photon flux units) through these operators as follows
\begin{eqnarray}\label{field}
E^{(+)}(t,z) &=& \int a(\Omega,z)e^{i\left( k_0z-(\omega_0+\Omega)t\right)}d\Omega\\\nonumber
             &=& \int b(\Omega,z)e^{i\left( k(\Omega)z-(\omega_0+\Omega)t\right)}d\Omega.
\end{eqnarray}

The operator $b\left(\Omega ,z\right)$ corresponds to the modal function, which is a solution of the wave equation in the absence of nonlinear interaction; therefore in its presence $b\left(\Omega ,z\right)$ is the slowly-varying amplitude. In terms of this operator the equation for the down-converted waves at frequencies $\omega_{\mathrm{0}}+\Omega $ and $\omega_{\mathrm{0}}-\Omega $ takes the well-known form \cite{Bloembergen65,Armstrong62},
\begin{equation}\label{waveeq}
\frac{\partial b(\Omega,z)}{\partial z}=\chi^{\left( 2
\right)}b_{p}b^{\dagger}(-\Omega,z)e^{i\Delta(\Omega)z},
\end{equation}
where $\chi^{\mathrm{(2)}}$ is the appropriately scaled element of the nonlinear susceptibility tensor of the second order, responsible for the nonlinear interaction, while $b_{\mathrm{p}}$ is the pump-wave amplitude in units of photon flux. An effective interaction of the tree waves is possible only for such frequencies $\Omega$ where the phase-matching condition $\Delta(\Omega)\approx 0$ is approximately satisfied. Usually, in an experiment the phase matching is realized for a narrow frequency band by selecting an angle of propagation with respect to the optical axis of the crystal \cite{Bloembergen65}, birefringence being taken into consideration.

If reaching the phase matching is impossible at the desired frequency $\omega_{\mathrm{0}}+\Omega $, one can apply the method of QPM, which consists of the following \cite{Armstrong62}. An artificial periodic layered structure is produced out of the original crystal, where the width of each layer is $\Lambda/2$, and each subsequent layer is different from the previous one by inversion of the crystal structure. As a result of such an inversion the second-order nonlinear susceptibility tensor changes its sign, though the linear properties of the crystal remain unchanged. The spatial modulation of the second-order nonlinear susceptibility in such a layered structure has the form of a meander
\begin{equation}\label{meander}
\chi^{(2)}(z)=\chi_{0}\sgn\left( \sin {Kz}
\right)=\frac{-i\chi_{0}}{\pi }\sum\limits_{n=-\infty }^{+\infty }
{\frac{1-\left( -1 \right)^{n}}{n}e^{inKz}},
\end{equation}
where $K=2\pi/\Lambda$ is the spatial frequency of the created grating, $\chi_{\mathrm{0}}$ is the second-order nonlinear susceptibility of the first layer, and the Fourier series decomposition of the meander function has been used, containing only odd values of $n$ (the term with $n=0$ is implied to be zero). Quasi-phase-matching of the first order for frequencies $\omega_{\mathrm{0}}+\Omega$ and $\omega_{\mathrm{0}}-\Omega $ consists of choosing the grating vector such that $K=\Delta(\Omega)$. In this case the additional phase factor, corresponding to $n=-1$, will compensate the phase mismatch at the desired frequency, when Eq.~(\ref{meander}) is substituted into Eq.~(\ref{waveeq}). All other terms in Eq.~(\ref{meander}) can be disregarded under typical conditions \cite{Powers11}.

In practice such periodically oriented crystals are created by a number of different methods \cite{Powers11,Houe95}. The most widely used of them is the method based on the property of a ferroelectric crystal to change its crystal structure under the action of an external electric field and then to maintain this structure when the external field is removed. Applying a spatially-periodic constant electric field to a ferroelectric with a significant second-order nonlinear susceptibility, such as lithium niobate, allows one to create artificial structures with QPM for practically any combination of wavelengths in various nonlinear optical processes. Such crystals are generally known as periodically poled and represent today a versatile tool in nonlinear optics.

\begin{figure}[htbp]
\includegraphics[width=\linewidth]{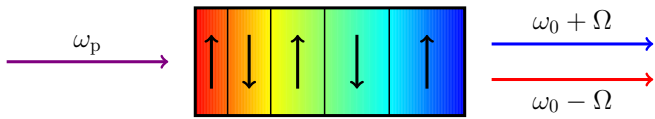}
\caption{\label{fig0} Parametric downconversion in an aperiodically poled crystal. The color varying along the crystal length shows the phase-matched signal frequency at the given position.}
\end{figure}

In the past decades much interest has been concentrated on the development of the above-described method, based on a slow change of the spatial frequency $K(z)$ along the crystal, allowing one to reach QPM at different frequencies in different parts of the crystal (Fig.~\ref{fig0}). Such crystals received the name of aperiodically poled crystals and are widely used for parametric amplification of ultrashort optical pulses \cite{Baker03,Fejer05,Fejer08a,Fejer08b,Heese10,Phillips14} and generation of broadband entangled photon pairs \cite{Harris07,Nasr08,Brida09,Sensarn10,Tanaka12}.

When the spatial frequency modulation is weak, the local period of the grating, $\Lambda (z)=2\pi /K(z)$, is a slowly varying function of the coordinate $z$ and under the condition $|\Lambda'(z)|\ll 1$ Eq.~(\ref{meander}) can be rewritten as
\begin{eqnarray}\label{meander2}
\chi^{(2)}(z)&=&\chi_{0}\sgn\left( \sin \int_0^z{K( z' )dz'} \right)\\\nonumber
&\approx& \frac{-i\chi_{0}}{\pi}\sum\limits_{n=-\infty }^{+\infty } {\frac{1-\left( -1
\right)^{n}}{n}e^{in\int_0^z {K\left( z' \right)dz'} }}.
\end{eqnarray}
Leaving only the term with $n=-1$ and substituting Eq.~(\ref{meander2}) into Eq.(\ref{waveeq}), we obtain
\begin{eqnarray}\label{waveeq2}
\frac{\partial b(\Omega ,z)}{\partial z}&=&i\gamma b^{\dagger}(-\Omega, z)e^{i\Delta(\Omega)z
-i\int_0^z {K( z' )dz'} },\\\nonumber
\frac{\partial b^{\dagger}(-\Omega ,z)}{\partial z}&=&-i\gamma^{\ast }b(\Omega,
z)e^{-i\Delta(\Omega)z+i\int_0^z {K( z' )dz'} },
\end{eqnarray}
where $\gamma = 2\chi_{\mathrm{0}}b_{\mathrm{p}}/\pi $ is the coefficient of nonlinear coupling, and the second equation is obtained from the first one by a Hermitian conjugation and a sign inversion for $\Omega $. We note, that the function $\Delta(\Omega)$ is even by definition for the considered case of type-I phase matching. Equations (\ref{waveeq2}) represent a closed system, having a unique solution for given boundary conditions. For finding this solution we introduce a new field operator $\tilde{b}(\Omega,z)$ by the following relation:
\begin{eqnarray}\label{btilde}
b(\Omega ,z)=\tilde{b}(\Omega,z)e^{ \frac{i}{2}\left(\Delta(\Omega)z-\int_0^z {K(z)dz} +\varphi_0\right)},
\end{eqnarray}
where $\varphi_{0}=\arg(i\gamma)$ combines the phases of the pump wave and $\chi_{\mathrm{0}}$. Now the system of Eqs. (\ref{waveeq2}) takes the form
\begin{eqnarray}\label{waveeq4}
\frac{\partial \tilde{b}(\Omega ,z)}{\partial z}&+&\frac{i}{2}\left( \Delta(\Omega)
-K(z) \right)\tilde{b}(\Omega, z)\\\nonumber
&=&|\gamma| \tilde{b}^{\dagger}(-\Omega, z)\\\nonumber
\frac{\partial \tilde{b}^{\dagger}(-\Omega ,z)}{\partial z}&-&\frac{i}{2}\left(\Delta(\Omega)
-K(z)\right)\tilde{b}^{\dagger}(-\Omega,z)\\\nonumber
&=&|\gamma|\tilde{b}(\Omega,z).
\end{eqnarray}
Solution of this system with the boundary conditions
$\tilde{b}(\Omega,0)$, $\tilde{b}^{\dagger}(-\Omega,0)$ will give a
transformation of the field operators in the nonlinear crystal. Practical
interest is represented by their values at the output of the crystal, at
the point $z=L$, where $L$ is the length of the crystal, i.e.,
$\tilde{b}(\Omega,L)$ and $\tilde{b}(-\Omega,L)$.

Excluding the operator $\tilde{b}^{\dagger}(-\Omega,z)$ from the system of Eqs. (\ref{waveeq4}), we obtain one equation of second order:
\begin{eqnarray}\label{waveeq6}
\frac{\partial^{2}\tilde{b}(\Omega ,z)}{{\partial z}^{2}}&+&\left(
\frac{1}{4}\left( \Delta(\Omega)-K( z )\right)^{2}-| \gamma |^{2}\right.\\\nonumber
&-&\left.\frac{i}{2}K'(z)\right)\tilde{b}(\Omega ,z)=0.
\end{eqnarray}
In the next section we discuss the general structure of the solution of this equation.

\subsection{The general structure of solution}

The system of Eqs.~(\ref{waveeq4}) with the boundary conditions at $z=0$ has a unique solution in the form of a Bogoliubov transformation for the field operators:
\begin{equation}\label{BogoliubovAB}
\tilde{b}(\Omega,L)=A(\Omega)\tilde{b}(\Omega,0)+B(\Omega)\tilde{b}^{\dagger}(-\Omega,0),
\end{equation}
where $A(\Omega)$ and $B(\Omega)$ are some complex functions. Note, that the frequency detuning enters Eqs.~(\ref{waveeq4}) only through $\Delta(\Omega)$ which is an even function. Therefore, the functions $A(\Omega)$ and $B(\Omega)$ are also even. Equation~(\ref{BogoliubovAB}) can be rewritten in terms of the sideband photons annihilation and creation operators as
\begin{equation}\label{BogoliubovUV}
a(\Omega,L)=U(\Omega)a(\Omega,0)+V(\Omega)a^{\dagger}(-\Omega,0),
\end{equation}
where
\begin{eqnarray}\label{U}
U(\Omega)&=&A(\Omega)e^{i[k(\Omega)-k_0+\frac{1}{2}\Delta(\Omega)]L-\frac{i}{2}\int_0^L{K(z)dz}},\\
\nonumber
V(\Omega)&=&B(\Omega)e^{i[k(\Omega)-k_0+\frac{1}{2}\Delta(\Omega)]L-\frac{i}{2}\int_0^L{K(z)dz}+i\varphi_{0}},
\end{eqnarray}
and these functions are not even in general because of their dependence on $k(\Omega)$.

The transformation (\ref{BogoliubovUV}) at a frequency where $V(\Omega)\ne 0$ corresponds to generation of a two-mode squeezed field state \cite{Kolobov99}. As any Bogoliubov transformation, it is fully characterized by four real parameters. Indeed, Eq.~(\ref{BogoliubovUV}) together with its Hermitian conjugate with opposite detuning represents a closed linear transformation for a pair of operators $\{a(\Omega,z),a^{\dagger}(-\Omega,z)\}$ from $z=0$ to $z=L$. This transformation for a fixed $\Omega$ is fully characterized by four complex numbers $U(\pm\Omega)$, $V(\pm\Omega)$. Unitarity of the Bogoliubov transformation imposes four real conditions $|U(\pm\Omega)|^2-|V(\pm\Omega)|^2=1$, and $U(\Omega)/V(\Omega)=U(-\Omega)/V(-\Omega)$ (note that the latter complex equation is equivalent to two real conditions), so only four real parameters remain. They can be defined as one squeezing parameter and three characteristic angles \cite{Kolobov99} by the following expressions:
\begin{eqnarray}\label{r}
r(\Omega)& = &\ln\left(|U(\Omega)|+|V(\Omega)|\right),\\\label{psiL}
\psi_L(\Omega)& = &\frac12 \arg\left[U(\Omega)V(-\Omega)\right],\\\label{psi0}
\psi_0(\Omega)& = &\frac12 \arg\left[U^{-1}(\Omega)V(\Omega)\right],\\\label{kappa}
\kappa(\Omega)& = &\frac12 \arg\left[U(\Omega)U^{-1}(-\Omega)\right],
\end{eqnarray}
where the first three parameters are even functions of $\Omega$, while the fourth one is odd.
The physical meaning of these parameters becomes clear from the definition of the squeezed quadrature. For each pair of modes with opposite detunings we construct two quadrature operators as~\cite{Kolobov99}
\begin{eqnarray}\label{quad1}
     X_1(\Omega,z)&=&a(\Omega,z)e^{-i\psi_z(\Omega)} +a^{\dagger}(-\Omega,z)e^{i\psi_z(\Omega)},\\
     \nonumber
     X_2(\Omega,z)&=&-i\left[a(\Omega,z)e^{-i\psi_z(\Omega)} -a^{\dagger}(-\Omega,z)e^{i\psi_z(\Omega)}\right],
\end{eqnarray}
where for the aims of the present discussion $z$ is equal only to zero and $L$. In terms of these quadratures the transformation Eq.~(\ref{BogoliubovUV}) can be rewritten in a simple form,
\begin{equation}\label{Xj}
     X_{j}(\Omega,L)=e^{\pm r(\Omega)+i\kappa(\Omega)}X_{j}(\Omega,0),
\end{equation}
where the upper (lower) sign corresponds to $j=1$ ($j=2$). It follows from Eq.~(\ref{Xj}) that the quadrature $X_2(\Omega,L)$ is squeezed below the standard quantum limit, while the conjugate quadrature $X_1(\Omega,L)$ is stretched above that limit. The squeezing parameter $r(\Omega)$ determines the degree of this effect, while the angle of squeezing, $\psi_L(\Omega)$, determines the quadrature at which the squeezing is to be observed at the output of the nonlinear crystal.

The angle $\psi_0(\Omega)$ determines the quadrature at the input, which is subject to the squeezing operation. For an unseeded PDC this angle is irrelevant, since all quadratures of the input field are in the vacuum state. However, for a seeded PDC this angle is to be taken into account, as discussed in Sec.~\ref{total}.

The last parameter $\kappa(\Omega)$ in our case of even $A(\Omega)$ and $B(\Omega)$ is independent of the nonlinear properties of the crystal and is given by
\begin{equation}\label{kappa}
    \kappa(\Omega)=\frac{1}{2}\left[(k(\Omega)-k(-\Omega)\right]L\approx \tau_g\Omega,
\end{equation}
where $\tau_g=k'(0)L$ is the time of light propagation through the crystal at the group velocity of the central wavelength of the downconverted light.

Below we are interested in finding the functions $r(\Omega)$, $\psi_L(\Omega)$, and $\psi_0(\Omega)$, characterizing the nonlinear transformation of the field in QPM crystals. In the next section we present an exact solution for a linear poling profile, while in Sec. IV we discuss in detail an approximate solution for a sufficiently slowly varying, but otherwise arbitrary, poling profile $K(z)$.

\section{Exact solution for a linear poling profile \label{linear}}

In this section we present an exact solution of the system of Eqs. (\ref{waveeq4}) in the case of linear chirp of the grating vector $K(z)=K_{0}-\zeta z$, where $\zeta>0$ is the chirp rate.

For a fixed $\Omega$ we introduce a new variable $x=\sqrt \zeta z+\left(\Delta(\Omega)-K_{0}\right)/\sqrt\zeta$. In the variables $(\Omega, x)$, Eqs.~(\ref{waveeq4}) for a linear chirp take the following form:
\begin{eqnarray}\label{wave1bis}
\frac{\partial \tilde{b}(\Omega,x)}{\partial
x}+\frac{i}{2}x\tilde{b}(\Omega,x)&=&\sigma \tilde{b}^{\dagger}(-\Omega,x),\\
\nonumber
\frac{\partial \tilde{b}^{\dagger}(-\Omega,x)}{\partial
x}-\frac{i}{2}x\tilde{b}^{\dagger}(-\Omega,x)&=&\sigma \tilde{b}(\Omega,x),
\end{eqnarray}
where $\sigma =|\gamma|/\sqrt\zeta$ is a new coupling
coefficient. The second-order equation (\ref{waveeq6}) in the new variables is
\begin{equation}\label{wave3bis}
\frac{\partial^{2}\tilde{b}(\Omega,x)}{{\partial x}^{2}}+\left(
\frac{1}{4}x^{2}-\sigma^{2}+\frac{i}{2}\right)\tilde{b}(\Omega,x)=0,
\end{equation}
and the corresponding equation for the operator $\tilde{b}^{\dagger}(-\Omega,x)$ is
\begin{equation}\label{wave4bis}
\frac{\partial^{2}\tilde{b}^{\dagger}(-\Omega,x)}{{\partial
x}^{2}}+\left(\frac{1}{4}x^{2}-\sigma^{2}-\frac{i}{2}\right)\tilde{b}^{\dagger}(-\Omega,x)=0.
\end{equation}

Equations (\ref{wave3bis}) and (\ref{wave4bis}) have solutions in the class of parabolic cylinder functions \cite{Abramowitz72}. Let us denote two linearly independent solutions of Eq.~(\ref{wave3bis}) with a constant Wronskian $W$ as $\phi_{1}(x)$ and $\phi_{2}(x)$. For these two functions we introduce ``reciprocal'' functions $\tilde{\phi}_{i}(x)$, $i=1,2$, by the relation
\begin{equation}\label{recip}
\frac{1}{\sigma}\left( \frac{\partial }{\partial x}+\frac{i}{2}x \right)\phi_{i}(x) =\tilde{\phi}_{i}(x).
\end{equation}

By construction, pairs $\left( \phi_{i}(x),\tilde{\phi}_{i}(x)\right)$ are solutions of the system of Eqs.~(\ref{wave1bis}). Let us prove that the functions $\tilde{\phi}_{1}(x)$ and $\tilde{\phi}_{2}(x)$ represent solutions of Eq.~(\ref{wave4bis}). This can be easily seen from writing Eqs.~(\ref{wave3bis}) and (\ref{wave4bis}) in an operator form, $T^{\ast}T\tilde{b}(\Omega,x) = \tilde{b}(\Omega,x)$ and $TT^{\ast}\tilde{b}^{\dagger}(-\Omega,x)=\tilde{b}^{\dagger}(-\Omega,x)$, respectively, where we have introduced a differential operator
\begin{equation}\label{T}
T=\frac{1}{\sigma}\left(\frac{\partial}{\partial x}+\frac{i}{2}x \right),
\end{equation}
having, after Eq.~(\ref{recip}), a meaning of mapping onto the ``reciprocal''
function: $T\phi_{i}(x)=\tilde{\phi }_{i}(x)$, and asterisk stands for complex conjugation. Substituting the last expression into the operator form of Eq.~(\ref{wave4bis}), and using the associative property of differential operators, we obtain $TT^{\ast}\tilde{\phi}_{i}(x)=(TT^{\ast})T\phi_{i}(x)=T(T^{\ast }T) \phi_{i}(x)=T\phi_{i}(x)=\tilde{\phi}_{i}(x)$, which had to be
proven. Also we easily obtain $T^{\ast}\tilde{\phi}_{i}(x)={T^{\ast}T\phi}_{i}(x)=\phi_{i}(x)$,
which is a complex conjugate operator $T^{\ast}$ that maps back the reciprocal
function onto the original one. Let us denote by $\tilde{W}$ the Wronskian
of functions $\tilde{\phi}_{1}(x)$ and $\tilde{\phi}_{2}(x)$. Then
\begin{eqnarray}\label{tildeW}
\tilde{W}&=&\left| \begin{array}{*{20}c}
\tilde{\phi}_{1}(x) & \tilde{\phi}_{2}(x)\\
\tilde{\phi}_{1}^{'}(x) & \tilde{\phi}_{2}^{'}(x)\\
\end{array} \right|
= \sigma \left| \begin{array}{*{20}c}
\tilde{\phi}_{1}(x) & \tilde{\phi}_{2}(x)\\
T^{\ast}\tilde{\phi}_{1}(x) & T^{\ast}\tilde{\phi}_{2}(x)\\
\end{array} \right|\\\nonumber
&=& \sigma \left| \begin{array}{*{20}c}
T\phi_{1}(x) & T\phi_{2}(x)\\
\phi_{1}(x) & \phi_{2}(x)\\
\end{array} \right|
= \left| \begin{array}{*{20}c}
\phi_{1}^{'}(x) & \phi_{2}^{'}(x)\\
\phi_{1}(x) & \phi_{2}(x)\\
\end{array} \right|=-W,
\end{eqnarray}
where we have used the property of invariance of the determinant under addition to one of its rows of another row, multiplied by an arbitrary factor. Equation~(\ref{tildeW}) shows that the reciprocal functions $\tilde{\phi}_{1}(x)$ and $\tilde{\phi}_{2}(x)$ are linearly independent if their original functions are.

Taking the complex conjugate of Eq.~(\ref{wave3bis}) in the operator form, we obtain $TT^{\ast}\phi_{i}^{\ast}(x)=\phi_{i}^{\ast}(x)$; i.e., the functions $\phi_{1}^{\ast}(x)$ and $\phi_{2}^{\ast}(x)$ are solutions of Eq.~(\ref{wave4bis}) and, therefore, are linear combinations of the functions $\tilde{\phi}_{1}(x)$ and $\tilde{\phi}_{2}(x)$. Let us write this dependence in the matrix form
\begin{equation}\label{matrix}
\left[ \begin{array}{*{20}c} \phi_{1}^{\ast}(x)\\
\phi_{2}^{\ast }(x)\\ \end{array} \right]
= \left[ \begin{array}{*{20}c} m_{11} & m_{12}\\
m_{21} & m_{22}\\ \end{array}  \right]
\left[ \begin{array}{*{20}c} \tilde{\phi }_{1}(x)\\
\tilde{\phi }_{2}(x)\\ \end{array} \right],
\end{equation}
where $m_{ij}$ are complex numbers. Now the Wronskian of $\phi_{1}^{\ast}(x)$ and $\phi_{2}^{\ast}(x)$ can be written as
\begin{eqnarray}\label{Wast}
W^{\ast}&=&\det \left[ \begin{array}{*{20}c}
\phi_{1}^{\ast}(x) & \phi_{2}^{\ast}(x)\\
\phi_{1}^{\ast '}(x) & \phi_{2}^{\ast '}(x)\\
\end{array} \right]\\\nonumber
&=& \det \left\{ \left[ \begin{array}{*{20}c}
\tilde{\phi}_{1}(x) & \tilde{\phi}_{2}(x)\\
\tilde{\phi}_{1}^{'}(x) & \tilde{\phi }_{2}^{'}(x)\\
\end{array} \right] M^{T} \right\}
=\tilde{W}\det M,
\end{eqnarray}
where $M$ is a matrix with the coefficients $m_{ij}$ from Eq.~(\ref{matrix}), and the superscript $T$ stands for transposition. Applying to Eq.~(\ref{matrix}) first the operator $T^{\ast}$, and then the complex conjugation, we obtain the property $M^{-1}=M^{\ast}$.

A general solution of the system of Eqs.~(\ref{wave1bis}) with the boundary conditions at the point $x_{0}=(\Delta(\Omega)-K_{0})/\sqrt\zeta$ can be written in the form
\begin{eqnarray}\label{AB}
\tilde{b}(\Omega,x)&=&A(x,x_{0})\tilde{b}(\Omega,x_{0})
+B(x,x_{0})\tilde{b}^{\dagger}(-\Omega ,x_{0}),\\
\nonumber
\tilde{b}^{\dagger}(-\Omega,x)&=&\tilde{A}(x,x_{0})\tilde{b}^{\dagger}(-\Omega ,x_{0})+\tilde{B}(x,x_{0})\tilde{b}(\Omega,x_{0}),
\end{eqnarray}
where
\begin{eqnarray}\label{A}
A(x,x_{0})&=&\frac{\sigma}{W}\left| \begin{array}{*{20}c}
\phi_{1}(x) & \phi_{2}(x)\\
\tilde{\phi}_{1}(x_{0}) & \tilde{\phi}_{2}(x_{0})\\
\end{array} \right|,\\\label{B}
B(x,x_{0})&=&-\frac{\sigma}{W}\left| \begin{array}{*{20}c}
\phi_{1}(x) & \phi_{2}(x)\\
\phi_{1}(x_{0}) & \phi_{2}(x_{0})\\
\end{array} \right|,\\\label{Atilde}
\tilde{A}(x,x_{0})&=&-\frac{\sigma}{W}\left| \begin{array}{*{20}c}\tilde{\phi}_{1}(x) & \tilde{\phi}_{2}(x)\\
\phi_{1}(x_{0}) & \phi_{2}(x_{0})\\
\end{array} \right|,\\\label{Btilde}
\tilde{B}(x,x_{0})&=&\frac{\sigma}{W}\left| \begin{array}{*{20}c}\tilde{\phi}_{1}(x) & \tilde{\phi}_{2}(x)\\
\tilde{\phi}_{1}(x_{0}) & \tilde{\phi}_{2}(x_{0})\\
\end{array} \right|.
\end{eqnarray}

The structure of the solution, Eqs.~(\ref{AB}), becomes more clear if we notice that both expressions are linear combinations of solutions of Eqs.~(\ref{wave3bis}) and (\ref{wave4bis}), respectively, and, therefore, are also their solutions. Moreover, $TA(x,x_{0})=\tilde{B}(x,x_{0})$ and $TB(x,x_{0})=\tilde{A}(x,x_{0})$, and therefore the pair of functions defined by Eqs.~(\ref{AB}) satisfies the system of Eqs.~(\ref{wave1bis}). Correspondence to the boundary conditions is seen from the following considerations. It is easy to see that $B(x_{0},x_{0})=\tilde{B}(x_{0},x_{0})=0$ because of the presence of two identical rows in both determinants. In addition,
\begin{eqnarray}\label{A00}
A(x_{0},x_{0})&=&\frac{\sigma }{W}\left| \begin{array}{*{20}c}
\phi_{1}(x_{0}) & \phi_{2}(x_{0})\\
\tilde{\phi}_{1}(x_{0}) & \tilde{\phi}_{2}(x_{0})\\
\end{array} \right|\\\nonumber
&=& \frac{\sigma}{W}\left| \begin{array}{*{20}c}
\phi_{1}(x_{0}) & \phi_{2}(x_{0})\\
T\phi_{1}(x_{0}) & T\phi_{2}(x_{0})\\
\end{array} \right|=1,
\end{eqnarray}
and similarly $\tilde{A}(x_{0},x_{0})=1$. Also, using Eqs.~(\ref{tildeW})-(\ref{Wast}), we find that
\begin{eqnarray}\label{Aast00}
A^{\ast}(x,x_{0})&=&\frac{\sigma}{W^{\ast}}\det\left\{
\left[ \begin{array}{*{20}c}
\tilde{\phi}_{1}(x) & \tilde{\phi}_{2}(x)\\
\phi_{1}(x_{0}) & \phi_{2}(x_{0})\\
\end{array} \right]M^{T} \right\} \\\nonumber
&=& -\frac{W}{W^{\ast}}\tilde{A}(
x,x_{0})\det M=\tilde{A}(x,x_{0}),
\end{eqnarray}
and similarly $B^{\ast}(x,x_{0})=\tilde{B}(x,x_{0})$; i.e., for the coefficients in Eqs.~(\ref{A})-(\ref{Btilde}) an exchange of the original and the reciprocal functions is equivalent to complex conjugation, though in general $\tilde{\phi}_{i}(x)\ne \phi_{i}^{\ast}(x)$. In particular, it follows that the second of Eqs.~(\ref{AB}) can be obtained from the first one by taking a Hermitian conjugation, as expected. It should be noted, that by definition $x$ is an even function of the frequency detuning $\Omega$, since it is determined by $\Delta(\Omega)$.

It is left to prove that Eqs.~(\ref{AB}), as required for a Bogoliubov transform, preserve the commutator of the field operators. To this end we need to show the fulfillment of two conditions \cite{Kolobov99}: evenness of $A(x,x_{0})/B(x,x_{0})$ as a function of $\Omega$ and the relation $\left|A(x,x_{0})\right|^2-\left|B(x,x_{0})\right|^{2}=1$. Fulfillment of the first condition follows from the evenness of both coefficients $A(x,x_{0})$ and $B(x,x_{0})$ as functions of $\Omega$. Let us show that the second condition is always satisfied by these coefficients:
\begin{eqnarray}\label{unity}
&&\left|A(x,x_{0})\right|^{2}-\left|B(x,x_{0})\right|^{2}\\\nonumber
&&= A(x,x_{0})\tilde{A}(x,x_{0})-B(x,x_{0})\tilde{B}(x,x_{0})\\\nonumber
&&= -\frac{\sigma^{2}}{W^{2}}\det \left\{ \left[ \begin{array}{*{20}c}
\phi_{1}(x) & \phi_{2}(x)\\
\tilde{\phi}_{1}(x_{0}) & \tilde{\phi}_{2}(x_{0})\\
\end{array} \right]
\left[ \begin{array}{*{20}c}\tilde{\phi}_{1}(x) & \phi_{1}(x_{0})\\
\tilde{\phi}_{2}(x) & \phi_{2}(x_{0})\\
\end{array} \right] \right\}\\\nonumber
&&+ \frac{\sigma^{2}}{W^{2}}\det \left\{ \left[
\begin{array}{*{20}c} \phi_{1}(x) & \phi_{2}(x)\\
\phi_{1}(x_{0}) & \phi_{2}(x_{0})\\
\end{array} \right] \left[ \begin{array}{*{20}c}
\tilde{\phi}_{1}(x) & \tilde{\phi}_{1}(x_{0})\\
\tilde{\phi}_{2}(x) & \tilde{\phi}_{2}(x_{0})\\
\end{array} \right] \right\}\\\nonumber
&&=\frac{\sigma^{2}}{W^{2}}\left|
\begin{array}{*{20}c}\phi_{1}(x_{0}) & \phi_{2}(x_{0})\\
\tilde{\phi}_{1}(x_{0}) & \tilde{\phi}_{2}(x_{0})\\
\end{array} \right|\left| \begin{array}{*{20}c}
\phi_{1}(x) & \tilde{\phi}_{1}(x)\\
\phi_{2}(x) & \tilde{\phi}_{2}(x)\\
\end{array} \right|=1,
\end{eqnarray}
where we have used the invariance of the matrix determinant to the operation of transposition.

Thus, Eqs.~(\ref{AB}) together with their Hermitian conjugation give a correct
description of the evolution of quantum field operators in the nonlinear
medium, in full correspondence to  Eq.~(\ref{BogoliubovAB}), with $A(\Omega)=A(x,x_{0})$ and $B(\Omega)=B(x,x_{0})$.

Let us consider Eq.~(\ref{BogoliubovUV}) for the sideband operator. The functions $U(\Omega)$ and $V(\Omega)$ are given by Eqs.~(\ref{U}), where the right-hand side is defined by the exact solution obtained above. We see easily that the transformation of Eq.~(\ref{BogoliubovUV}) is unitary. The ratio $U(\Omega)/V(\Omega)$ is even as a function of $\Omega$ because it is proportional to $A(\Omega)/B(\Omega)$, which is even. The relation $\left|U(\Omega)\right|^2-\left|V(\Omega)\right|^{2}=1$ follows from the corresponding properties of the functions $A(\Omega)$ and $B(\Omega)$, since these functions differ only by phase.

For practical calculations in the rest of this article we choose the
functions $\phi_{1}(x)$ and $\phi_{2}(x)$ from the family of Whittaker
functions (see \S 19.3.7 in Ref.~\cite{Abramowitz72}), which are represented in the system of computer algebra
Mathematica 10. Thus, for a fixed parameter $\nu =\sigma^2$ we let
\begin{eqnarray}\label{phi1}
\phi_{1}(x)&=&D_{i\nu}(xe^{i\pi/4}),\\
\nonumber
\phi_{2}(x)&=&D_{-1-i\nu }(-xe^{-i\pi/4}),
\end{eqnarray}
with the corresponding reciprocal functions
\begin{eqnarray}\label{phitilde1}
\tilde{\phi}_{1}(x)&=&{\nu^{1/2}e}^{i3\pi/4}D_{i\nu-1}(xe^{i\pi/4}),\\
\nonumber
\tilde{\phi}_{2}(x)&=&{\nu^{-1/2}e}^{-i\pi/4}D_{-i\nu}(-xe^{-i\pi/4}).
\end{eqnarray}
The Wronskian of the functions, defined by Eqs.~(\ref{phitilde1}), is equal to $W=e^{-i\pi/4}e^{\pi\nu/2}$. The spectra of PDC calculated with these functions are presented in Sec.~\ref{spectra}.

Thus, we have seen that in the case of a linear poling profile an analytic solution for the Heisenberg equations of motion for the field exists in the class of special functions. Unfortunately, in the general case of the nonlinear profile it is not so, and the solution can be computed numerically only. In the next section we show how an approximate solution can be obtained in a more general case.

\section{Approximate solution for a nonlinear poling profile \label{approx}}

\subsection{Formulation of the equivalent ``potential barrier'' problem}

In this section we derive an approximate solution for the field transformation in an aperiodically poled crystal in a very simple analytic form. Our approach is based on the similitude of the field evolution to that of a quantum particle in a given potential and is similar to the approach of Refs.~\cite{Baker03,Fejer08a} with the main difference that we consider the evolution of a quantum field and are interested in a unitary transformation of the field operators. In addition, we obtain the approximate solution directly in the first-order approximation, without deriving a second-order solution and then simplifying it, as in Ref.~\cite{Fejer08a}.

Equation (\ref{waveeq6}) is similar to the Schr\"odinger equation for a particle of mass $1/2$ in a given potential,
\begin{eqnarray}\label{Schrodinger}
\frac{\partial^2}{\partial z^2}\Psi(z)+\left(E-\mathcal{U}(z)\right)\Psi(z)=0,
\end{eqnarray}
where $\Psi(z)$ is the particle wavefunction, $E=-|\gamma|^{2}$ is the energy of the particle, and the potential is defined as
\begin{equation}\label{Uz}
\mathcal{U}(z)=-\frac{1}{4}\left(\Delta(\Omega)-K(z) \right)^{2}.
\end{equation}
Rewriting Eq.~(\ref{waveeq6}) in the form of Eq.~(\ref{Schrodinger}), we have omitted the term $K'(z)$, which is justified for a sufficiently slowly varying profile \cite{Baker03,Fejer08a}.

An approximate solution of Eq.~(\ref{Schrodinger}) can be obtained in the first-order approximation \cite{Landau65}. In this approximation the solution is oscillating in the regions where $\mathcal{U}(z)<E$ and exponentially growing or decaying in the regions where $\mathcal{U}(z)>E$. For the sake of simplicity we limit ourselves to monotonous profiles $K(z)$, which, for definiteness, we consider to be decreasing functions of $z$.

\begin{figure}[htbp]
\includegraphics[width=\linewidth]{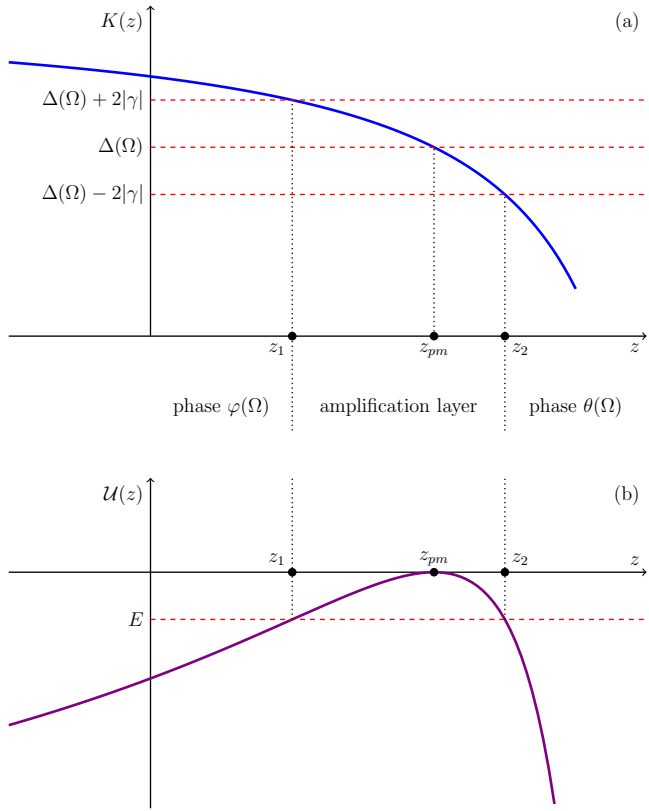}
\caption{\label{Diagram} Three regions in the crystal for a given detuning $\Omega$. (a) Spatial frequency of aperiodic poling as a function of position in the crystal. (b) Potential for an equivalent problem for a quantum particle of energy $E$, passing through a potential barrier, solved semiclassically with both exponentially growing and decaying solutions. The point of perfect phase match $z_{pm}$ corresponds to the peak of the barrier, and the turning-points $z_{1}$ and $z_{2}$ correspond to reflection of a classical particle coming to $z_{1}$ from the left or to $z_{2}$ from the right.}
\end{figure}

In a crystal with a monotonous profile $K(z)$ for every pair of frequencies $\omega_{0}+\Omega $ and $\omega_{0}-\Omega $ from the parametric amplification band, there is a perfect phase-matching point $0\le z_{pm}(\Omega)\le L$, defined by the relation
\begin{equation}\label{zpm}
K\left( z_{pm}(\Omega ) \right)=\Delta(\Omega).
\end{equation}
At this point the potential $\mathcal{U}(z)$ is maximal and equal to zero. To the left and right of this point there are the so-called ``turning points'', where $\mathcal{U}(z)=E$. These points are defined by the relation
\begin{equation}\label{tp}
K\left( z_{1,2}(\Omega ) \right)=\Delta\left(\Omega\right)\pm 2|\gamma|
\end{equation}
and represent the borders of the region of the exponential solution (see Fig.~\ref{Diagram}). Note that in our case the oscillatory solutions exist in the regions $(-\infty,z_{1}]$ and $[z_{2},\infty)$, while the exponentially growing and decaying solutions, corresponding to the parametric amplification and attenuation, exist in the region $[z_{1},z_{2}]$. Therefore, our equivalent quantum particle problem corresponds to passing through a ``potential barrier'' and not to oscillating in a ``potential well.''

Using the approach described above, the evolution of the signal field in the crystal can be represented by the following simplified picture. In the region $[0,z_{1}(\Omega)]$ the field remains in its vacuum state. A major part of photons at a given pair of frequencies $\omega_{0}+\Omega $ and $\omega_{0}-\Omega $ is generated in a narrow layer of crystal between $z_{1}(\Omega)$ and $z_{2}(\Omega)$, which we call the amplification layer. Afterwards both waves propagate through the crystal with practically unchanging amplitudes, acquiring a phase difference due to the crystal dispersion. The field operator at the crystal output $\tilde{b}(\Omega ,L)=\Psi(L)$ is given by the solution of Eq.~(\ref{Schrodinger}) with the conditions at some point $z=z_0$,
\begin{eqnarray}\label{initial1}
\Psi(z_0)&=&\tilde{b}(\Omega ,z_0),\\
\nonumber
\Psi'(z_0)&=&\pm i\sqrt{-\mathcal{U}(z_0)}\tilde{b}(\Omega ,z_0)+\sqrt{-E} \tilde{b}^{\dagger}(-\Omega, z_0).
\end{eqnarray}
Here the second condition is derived from Eq.~(\ref{waveeq4}) and for a decreasing poling profile the upper (lower) sign should be taken for $z_0<z_{pm}$ ($z_0>z_{pm}$). Note that in the equivalent ``potential barrier'' problem $\Psi(z)$ is considered as a c-number. However, after a solution of Eq.~(\ref{Schrodinger}) is obtained in the form of a linear combination of initial values $\Psi(0)$ and $\Psi'(0)$, the latter can be substituted by operator-valued expressions, Eqs.~(\ref{initial1}) with $z_0=0$, corresponding to the original physical problem.

In what follows we derive a solution of Eq.~(\ref{Schrodinger}) together with the conditions in Eqs.~(\ref{initial1}) in the first-order approximation. A second-order, Wentzel-Kramers-Brillouin (WKB) solution, was obtained for a classical field in the high-gain regime in Refs.~\cite{Baker03,Fejer08a}. However, below we demonstrate that even the first-order solution describes very well the average shapes of optical and squeezing spectra and with a very high precision the characteristic squeezing angles. In the calculations which follow we often omit the frequency detuning, which is always equal to $\Omega$.

\subsection{Oscillating solution before amplification \label{before}}

In the first-order approximation the solution of Eq.~(\ref{Schrodinger}) in the region $[0,z_{tp1}]$ is given by \cite{Landau65}
\begin{equation}\label{Psi1}
\Psi(z)=C_{1}e^{+i\int_0^{z}{\sqrt {E-\mathcal{U}(z)}dz}}+C_{2}e^{-i\int_0^{z}{\sqrt {E-\mathcal{U}(z)}dz}},
\end{equation}
where $C_1$ and $C_2$ are constants, to be determined from the initial conditions. In the considered region almost everywhere we have $|E|\ll|\mathcal{U}(z)|$. Thus, disregarding $E$ compared to $\mathcal{U}(z)$ in Eqs.~(\ref{initial1}) and (\ref{Psi1}), we obtain $C_2=0$ and write the solution in the form of a phase shift $\tilde{b}(\Omega,z_{1})=\tilde{b}(\Omega,0)e^{i\varphi (\Omega)}$, with
\begin{equation}\label{varphi}
\varphi (\Omega)=-\frac{1}{2}\int_0^{z_{pm}\left( \Omega \right)} {\left( \Delta(\Omega)-K(z) \right)dz},
\end{equation}
where we have replaced the upper integration limit $z_{1}(\Omega)$ by $z_{pm}(\Omega)$, which is a good approximation for a sufficiently thin amplification layer.

\subsection{Exponential solution inside the amplification layer}

In the first-order approximation the solution of Eq.~(\ref{Schrodinger}) in the region $[z_{1},z_{2}]$ is given by \cite{Landau65},
\begin{equation}\label{Psi2}
\Psi(z)=\tilde C_{1}e^{+\int_{z_1}^{z}{\sqrt {\mathcal{U}(z)-E}dz}}+\tilde C_{2}e^{-\int_{z_1}^{z}{\sqrt {\mathcal{U}(z)-E}dz}},
\end{equation}
where $\tilde C_1$ and $\tilde C_2$ are some constants. Unfortunately, these constants cannot be obtained from the initial condition at $z=z_1$ since it is a turning point, where $\mathcal{U}(z_1)=E$ and therefore $\Psi'(z_1)=0$. The problem of tailoring the solutions at the turning points is well known for both the first-order and the WKB approximations \cite{Landau65}. Below we show how this problem can be circumvented in our case.

Accepting that the amplification layer is very thin compared to the distance at which the poling profile $K(z)$ is substantially nonlinear, we can approximate the profile inside the amplification layer by its Taylor expansion around $z=z_{pm}$ up to the linear term: $K(z)\approx K(z_{pm})+K'(z_{pm})(z-z_{pm})$. Substituting such a linearized profile into Eq.~(\ref{Psi2}) and performing the integration, we obtain
\begin{equation}\label{Psi2bis}
\Psi(z)=\tilde C_{1}e^{\nu(\xi_z+\frac12\sin{2\xi_z}+\frac\pi2)}+\tilde C_{2}e^{-\nu(\xi_z+\frac12\sin{2\xi_z}+\frac\pi2)},
\end{equation}
where $\xi_z=\arcsin{s_z}$, and $s_z=|K'(z_{pm})|(z-z_{pm})/(2|\gamma|)$ is the normalized coordinate inside the amplification layer varying from $-1$ to $1$. Note that the parameter $\nu=|\gamma^2/K'(z_{pm})|$ is defined exactly as in Sec.~\ref{linear}, if we replace $\zeta$ by the local chirp rate $|K'(z_{pm})|$.

Substituting $z=z_2$ into Eq.~(\ref{Psi2bis}) gives the following expression for the field at the crystal output,
\begin{equation}\label{Psi2out}
\tilde{b}(\Omega,z_2)=\tilde C_1e^{\pi\nu}+\tilde C_2e^{-\pi\nu}.
\end{equation}
Taking the condition in Eqs.~(\ref{initial1}) at $z_0=z_1$ we obtain $\tilde C_{1}+\tilde C_{2}=\tilde{b}(\Omega,z_{1})$. In the absence of the second initial condition, Eq.~(\ref{Psi2out}) represents a family of solutions, from which one member should be selected with some considerations. Let us parametrize properly this family. The solution should be a linear combination of $\tilde{b}(\Omega,z_1)$ and $\tilde{b}^\dagger(-\Omega,z_1)$. Let us write
\begin{eqnarray}\label{C1}
\tilde C_1 &=& \frac{1+\mu}2\tilde{b}(\Omega,z_1)+\frac{\tilde\mu}2\tilde{b}^\dagger(-\Omega,z_1),\\
\nonumber
\tilde C_2 &=& \frac{1-\mu}2\tilde{b}(\Omega,z_1)-\frac{\tilde\mu}2\tilde{b}^\dagger(-\Omega,z_1),
\end{eqnarray}
where $\mu$ and $\tilde\mu$ are some complex coefficients. Such a parametrization is the most general one satisfying the relation for the sum of $\tilde C_1$ and $\tilde C_2$. Now Eq.~(\ref{Psi2out}) has the form
\begin{eqnarray}\label{Psi2ter}
\tilde{b}(\Omega,z_2) &=& \left[\cosh{(\pi\nu)}+\mu\sinh{(\pi\nu)}\right]\tilde{b}(\Omega,z_1)\\\nonumber
&+& \tilde\mu\sinh{(\pi\nu)}\tilde{b}^\dagger(-\Omega,z_1).
\end{eqnarray}
Unitarity of this transformation demands
\begin{equation}\label{beta}
|\tilde\mu| = \frac{\sqrt{\left|\cosh{(\pi\nu)}+\mu\sinh{(\pi\nu)}\right|^2-1}}{\sinh{(\pi\nu)}}.
\end{equation}

There are two candidates for $\mu$, met in similar physical problems. First, we notice that the case $\mu=0$, $|\tilde\mu|=1$ resembles the field transformation in a medium with perfect phase matching \cite{Kolobov99}. Indeed, in this case the signal field is multiplied (up to a phase) by $\cosh(gl)$, where $g$ is proportional to the pump amplitude and $l$ is the length of the medium. In our case the width of the medium (amplification layer) is also proportional to the pump amplitude [see Eq.~(\ref{tp})] and, as consequence, $gl$ is proportional to the pump intensity, exactly as the parameter $\nu$. However, such a choice is related to neglecting the phase mismatch close to the edges of the amplification layer, which may become significant with growing pump power, leading to widening of the amplification layer. Second, the case of $\mu=1$ resembles a solution obtained by Rosenbluth for a similar problem in plasma physics. In Ref.~\cite{Rosenbluth72} a parametric interaction of three waves in plasma is considered, which is governed by Eq.~(\ref{waveeq6}), written for c-numbers. In our case c-numbers appear when one considers the mean field,  $\langle \tilde{b}(\Omega ,z) \rangle$, which, of course, satisfies the same Eq.~(\ref{waveeq6}) because of its linearity. The initial conditions of Ref.~\cite{Rosenbluth72} correspond to the presence of the mean field at the input at the signal frequency, but not at the idler one, which is also a typical scenario of parametric amplification in classical optics \cite{Baker03,Fejer05,Fejer08a,Fejer08b,Heese10}. Rosenbluth's solution \cite{Rosenbluth72} can be written for the mean field as $\langle \tilde{b}(\Omega ,z_{2}) \rangle = \langle \tilde{b}(\Omega ,z_{1}) \rangle e^{\pi \nu}$, which obviously corresponds to $\mu=1$ in Eq.~(\ref{Psi2ter}). Unfortunately, there is no clear intuitive reason for giving a preference to $\mu=0$ or $\mu=1$, or maybe some other value of $\mu$.

\begin{figure}[htbp]
\includegraphics[width=\linewidth]{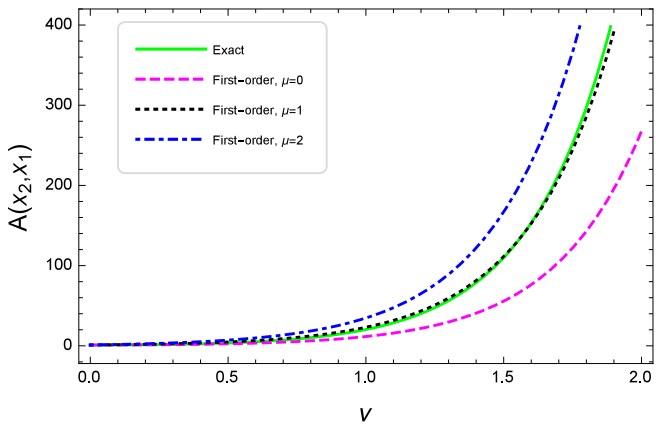}
\caption{\label{Fig_Mu} Amplitude gain at given frequency as function of the normalized pump power. Solid green line, exact solution for the linearized profile. Dotted, dashed, and dot-dashed lines represent approximate solutions for different values of the parameter $\mu$.}
\end{figure}

Fortunately, the field transformation for a linearized poling profile can be written in an analytic form by the approach of the previous section where the constant chirp rate $\zeta$ is to be substituted by the local chirp rate $|K'(z_{pm})|$ and $K_0$ by $K(z_{pm})-K'(z_{pm})z_{pm}$. Thus, the exact form of the field transformation in the amplification layer for a linearized profile, but without first-order approximation for an equivalent problem, is given by Eq.~(\ref{AB}) with $x=\sqrt\zeta(z-z_{pm})$. It is easy to find that the turning points correspond to the values $x_{1,2}=\pm2\sqrt\nu$. Equation~(\ref{A}) in our case takes the form
\begin{eqnarray}\label{Aal}
A(x_2,x_1) &=& e^{-\pi\nu/2}\left(\left|D_{i\nu}(2\sqrt\nu e^{i\pi/4})\right|^2\right.\\\nonumber
&+& \left.\nu \left|D_{i\nu-1}(-2\sqrt\nu e^{i\pi/4})\right|^2\right),
\end{eqnarray}
where we have used the expression of the elementary solutions through the Whittaker functions given by Eqs.~(\ref{phi1}) and (\ref{phitilde1}). We see that $A(x_2,x_1)$ is real, which corresponds to the choice of a real $\mu$. In Fig.~\ref{Fig_Mu} we show the amplitude gain as a function of $\nu$ for the exact solution and the approximate solutions corresponding to different values of $\mu$. We see that in the region of significant, but not too high squeezing, $0.5<\nu<2$, the case of $\mu=1$, i.e., the Rosenbluth formula, complies very well with the exact solution. A similar analysis of $B(x_2,x_1)$ shows that its phase is a slow function of $\nu$ and for $\nu<2$ can be approximated as $\varphi_1=\arg[B(x_2,x_1)]\approx -\nu+\nu^2/4$.

Thus, the total transformation in the amplification layer can be written as
\begin{equation}\label{Bogol}
\tilde{b}( \Omega ,z_{2} )=e^{\pi \nu(\Omega)} \tilde{b}(\Omega,z_{1})+e^{i\varphi_1}\sqrt{e^{2\pi\nu(\Omega)}-1}\tilde{b}^{\dagger}(-\Omega,z_{1}),
\end{equation}
which can be viewed as a quantum extension of the Rosenbluth formula.

\subsection{Oscillating solution after amplification}

In the region $[z_{2},L]$, analogously to Sec.~\ref{before}, the solution of Eq.~(\ref{Schrodinger}) in the first-order approximation is given by
\begin{equation}\label{Psi3}
\Psi(z)=\bar C_{1}e^{+i\int_{z_2}^{z}{\sqrt {E-\mathcal{U}(z)}dz}}+\bar C_{2}e^{-i\int_{z_2}^{z}{\sqrt {E-\mathcal{U}(z)}dz}},
\end{equation}
where $\bar C_1$ and $\bar C_2$ are constants, to be determined from the initial conditions. In the considered region again, almost everywhere we have $|E|\ll|\mathcal{U}(z)|$, and disregarding $E$ compared to $\mathcal{U}(z)$ in Eqs.~(\ref{initial1}) and (\ref{Psi3}), we obtain $\bar C_2=0$ and write the solution in the form of a phase shift $\tilde{b}(\Omega,L)=\tilde{b}(\Omega,z_2)e^{i\theta (\Omega)}$, with
\begin{equation}\label{theta}
\theta(\Omega)=-\frac{1}{2}\int_{z_{pm}\left( \Omega \right)}^L {\left( \Delta(\Omega)-K(z) \right)dz},
\end{equation}
where we have replaced the integration limit $z_{2}(\Omega)$ by $z_{pm}(\Omega)$.

\subsection{Total trasformation and the characteristic angles \label{total}}

Combining the results of the previous three sections and coming back to the sideband operator $a(\Omega,z)$, we write the total field transformation in the crystal in the following form:
\begin{equation}\label{Bogol2}
a(\Omega,L)=U_1(\Omega)a(\Omega,0)+V_1(\Omega)a^{\dagger}(-\Omega,0),
\end{equation}
where
\begin{eqnarray}\label{U1}
U_1(\Omega)&=&e^{\pi \nu (\Omega)}e^{i(k(\Omega)-k_0)L},\\
\nonumber
V_1(\Omega)&=&\sqrt{e^{2\pi\nu(\Omega)}-1}e^{-2i\varphi(\Omega)+i(k(\Omega)-k_0)L+i\varphi_A}.
\end{eqnarray}
Here index $1$ stands for the first-order approximation, and $\varphi_A=\varphi_0+\varphi_1$ is the phase added in the amplification layer. The properties $|U_1(\pm\Omega)|^2-|V_1(\pm\Omega)|^2=1$ and
$U_1(\Omega)/V_1(\Omega)=U_1(-\Omega)/V_1(-\Omega)$, required for the unitarity of this transformation, are straightforward to verify.

We show in the next section that the first-order approximate solution, given by Eq.~(\ref{Bogol2}) is very close to the analytical and numerical solutions of the initial Heisenberg equation for both small and considerable values of the pump power. In this way it differs from the solution, called the ``Rosenbluth formula'' in Ref.~\cite{Fejer08a}, which provides the same value $e^{\pi \nu}$ for moduli of  $U_1(\Omega)$ and $V_1(\Omega)$ inside the amplification band and is valid only at $\pi\nu\gg 1$. In this limit our Eqs.~(\ref{U1}) give the same result. At the same time, they give also a very good approximation at a moderate pump power, where $\pi\nu$ is less than or comparable to unity.

The three parameters in Eqs.~(\ref{r})-(\ref{psi0}), characterizing the nonlinear transformation of the field in the first-order approximation, are equal to
\begin{eqnarray}\label{r1}
r(\Omega)      & = &\ln\left(e^{\pi \nu (\Omega)}+\sqrt{e^{2\pi\nu(\Omega)}-1}\right),\\\label{psiL1}
\psi_L(\Omega) & = &-\varphi(\Omega)-\frac{1}{2}\Delta(\Omega)L+\frac{\varphi_A}2,\\\label{psi01}
\psi_0(\Omega) & = &-\varphi(\Omega)+\frac{\varphi_A}2.
\end{eqnarray}

As mentioned in Sec.~II, the angle $\psi_0(\Omega)$ determines the quadrature at the input which is subject to the squeezing effect and is important in the case of seeded PDC. For example, to obtain an amplitude squeezing at all frequencies, one needs to shape the signal seed pulse so that $\langle X_1(\Omega,0)\rangle=0$, $\langle X_2(\Omega,0)\rangle\ne0$, where at each frequency the quadratures are determined by $\psi_0(\Omega)$. For this the idler component should be equal to the conjugated and phase-shifted signal component, $\langle a(-\Omega,0)\rangle=-e^{i2\psi_0(\Omega)}\langle a(\Omega,0)\rangle^*$, where we have assumed $\Omega>0$. Equation~(\ref{psi0}) shows that this phase shift is exactly compensated by the relative phase difference $-2\varphi(\Omega)$ acquired by the idler field with respect to the signal field before the amplification layer, and by the phase $\varphi_A$, added during the amplification. As a result, the squeezing is in-phase with the coherent component of the field, as required for the amplitude squeezing.

Now we proceed to explain the physical meaning of the expression for the angle of squeezing, Eq.~(\ref{psiL}). Differentiating this equation and taking into account Eq.~(\ref{varphi}), we obtain
\begin{equation}\label{psidiff}
\frac{d\psi_L(\Omega)}{d\Omega} = -\frac{1}{2}\Delta'(\Omega)\left[ L-z_{pm}(\Omega) \right].
\end{equation}

Equation (\ref{psidiff}) has a simple physical meaning in terms of the classical notion of the relative group delay between the signal and the idler waves. In the classical treatment of PDC the field operators $a(\pm\Omega,z)$ are replaced by classical complex amplitudes $\langle a(\pm\Omega,z)\rangle$ ($\Omega$ is assumed to be positive). Equation (\ref{Bogol2}) with the corresponding replacement gives the field amplitudes at the crystal output. As mentioned above, in the scenario of parametric amplification, treated classically, it is typically assumed that at the crystal input only the signal wave is present, i.e., $\langle a(\Omega,0)\rangle=1$, $\langle a(-\Omega,0)\rangle=0$. In this case the signal wave at the output is equal to $\langle a(\Omega,L)\rangle=U(\Omega)$ and the idler wave to $\langle a(-\Omega,L)\rangle=V(-\Omega)$. Then, the group delay of the signal wave is given by $\tau_s(\Omega)=-\frac{d}{d\Omega}\arg\{U(\Omega)\}$ and that of the corresponding idler wave by $\tau_i(\Omega)=-\frac{d}{d(-\Omega)}\arg\{V(-\Omega)\}$. The relative delay is
\begin{eqnarray}\label{ang}
\tau(\Omega)&=&\tau_s(\Omega)-\tau_i(\Omega) \\\nonumber
&=&-\frac{d}{d\Omega}\arg\{U(\Omega)V(-\Omega)\}=-2\frac{d\psi_L(\Omega)}{d\Omega}.
\end{eqnarray}

Now Eq.~(\ref{psidiff}) can be rewritten as
\begin{equation}\label{taurel}
\tau(\Omega)=\frac{L-z_{pm}(\Omega)}{v_g(\Omega)}-\frac{L-z_{pm}(\Omega)}{v_g(-\Omega)},
\end{equation}
where $v_g(\Omega)=[k'(\Omega)]^{-1}$ is the group velocity at the corresponding frequency. Equation (\ref{taurel}) shows that the relative delay is equal to the difference of propagation times of two waves from the perfect phase-matching point to the end of the crystal with the corresponding group velocities, which is quite a natural result. This result is well known in the classical consideration of parametric amplification; see, e.g., Ref.~\cite{Fejer05}. In our quantum treatment we have shown that the relative group delay is related to the angle of squeezing via Eq.~(\ref{ang}).

Before finishing this section, let us estimate the condition of ``slow'' variation for the poling profile. In order that its linearization inside the amplification layer be valid, it is necessary that the second-order term in the Taylor expansion is much smaller than the first-order term, i.e., the parameter
\begin{equation}\label{epsilon}
\epsilon=\frac12\frac{|K''(z_{pm})(z_{tp}-z_{pm})|}{|K'(z_{pm})|}
\end{equation}
 should be much smaller than unity, where $z_{tp}$ is the most distant of two turning-points with respect to $z_{pm}$. We may obtain a more compact expression for the smallness parameter. If $\epsilon\ll1$ and the linearization is valid, then $|z_{tp}-z_{pm}|\approx 2|\gamma/K'(z_{pm})|$ and the parameter
\begin{equation}\label{epsilonprime}
\epsilon'=\frac{|\gamma K''(z_{pm})|}{(K'(z_{pm}))^2}
\end{equation}
is also much smaller than unity. Thus, $\epsilon'\ll1$ is a necessary condition for the applicability of the first-order approximation of this section. Note that, since $|\gamma|$ is proportional to $|b_{\mathrm{p}}|$, this condition imposes a limitation on the pump power.

\section{Spectra of parametric down-conversion for a model crystal \label{spectrasection}}
\subsection{Linearly chirped crystal: comparison of exact and approximate solutions \label{spectra}}

\begin{figure}[htbp]
\centerline{\includegraphics{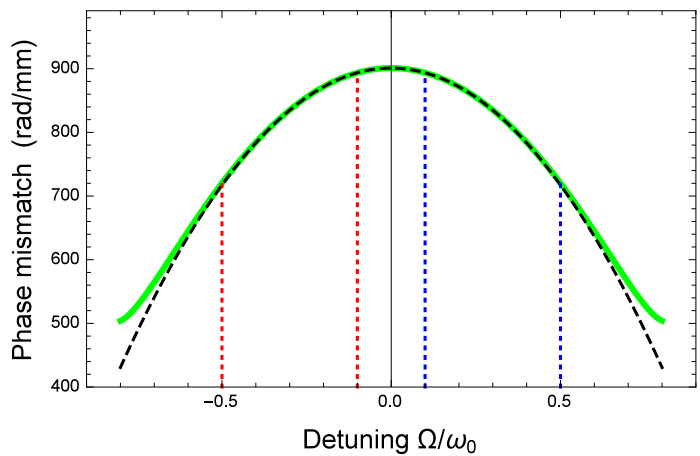}}
\caption{\label{fig1} Phase mismatch of a crystal of 5\% MgO-doped congruent LiNbO$_3$, pumped at 532 nm. Solid green line is obtained from the Sellmeier equation \cite{Gayer08}; dashed black line is its quadratic approximation. Dotted lines show the limits of spectral bands for the signal (blue) and the idler (red) waves. Both waves, as well as the pump wave, are extraordinary.}
\end{figure}

For making a comparison of the exact and the approximate solutions of the
wave equation we consider PDC in a nonlinear crystal of 5\% MgO-doped congruent aperiodically poled LiNbO$_3$ of length $L=$4.5 mm, continuously pumped at the wavelength $\lambda_p=532$ nm. The pump frequency is $\omega_{\mathrm{p}}=2\pi c/\lambda_p$, and the central frequency of the downconverted light is $\omega_0=\omega_{\mathrm{p}}/2$. The signal band is chosen to be from $1.1\omega_0$ to $1.5\omega_0$, and the idler band from $0.5\omega_0$ to $0.9\omega_0$. This corresponds to signal wavelengths of 709--967 nm, and to idler wavelengths of 1182--2128 nm. To obtain the desired frequency band of the downconverted light we need to vary the spatial frequency of aperiodical poling $K(z)$ from $K_0=\Delta(0.1\omega_0)=894$ rad/mm to $K_1=\Delta(0.5\omega_0)=720$ rad/mm. In this subsection we consider a  linear dependence $K_L(z)=K_0-\zeta z$, where $\zeta=(K_0-K_1)/L=38.5$ rad/mm$^2$. The phase mismatch for such a crystal, obtained from its Sellmeier equation \cite{Gayer08}, is shown in Fig.~\ref{fig1}. In the same figure we show the quadratic approximation of the phase mismatch
\begin{equation}\label{Dkq}
\Delta_q(\Omega)=-\alpha\left(\frac{\Omega}{\omega_0}\right)^2+\beta,
\end{equation}
where $\alpha=-\Delta''(0)\omega_0^2/2=735$ rad/mm, and $\beta=\Delta(0)=901$ rad/mm.

The optical spectrum of PDC, $S(\omega)$, is defined by the relation
$\langle a^\dagger(\Omega,L)a(\Omega',L)\rangle=S(\omega_{0}+\Omega)\delta(\Omega -\Omega')$, wherefrom, taking into account the commutation relations $\left[a(\Omega,z),a^{\dagger}(\Omega',z)\right]=\frac{1}{2\pi}\delta \left(\Omega -\Omega'\right)$ and Eq. (\ref{BogoliubovUV}), we obtain
\begin{equation}\label{S}
S(\omega_{0}+\Omega)=\frac{1}{2\pi }\left|V(\Omega)\right|^2.
\end{equation}

\begin{figure}[htbp]
\subfloat{\includegraphics[width=.48\linewidth]{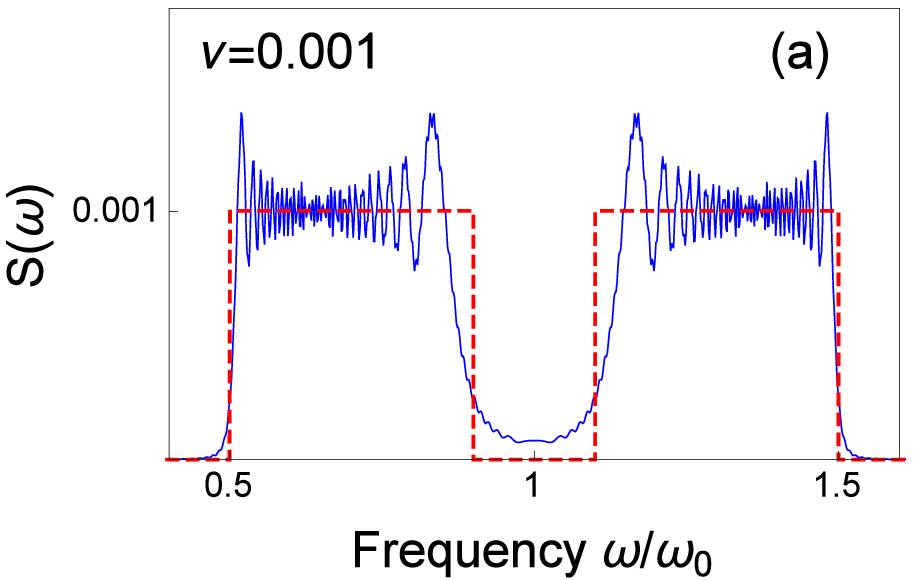}}\hspace{.02\linewidth}
\subfloat{\includegraphics[width=.48\linewidth]{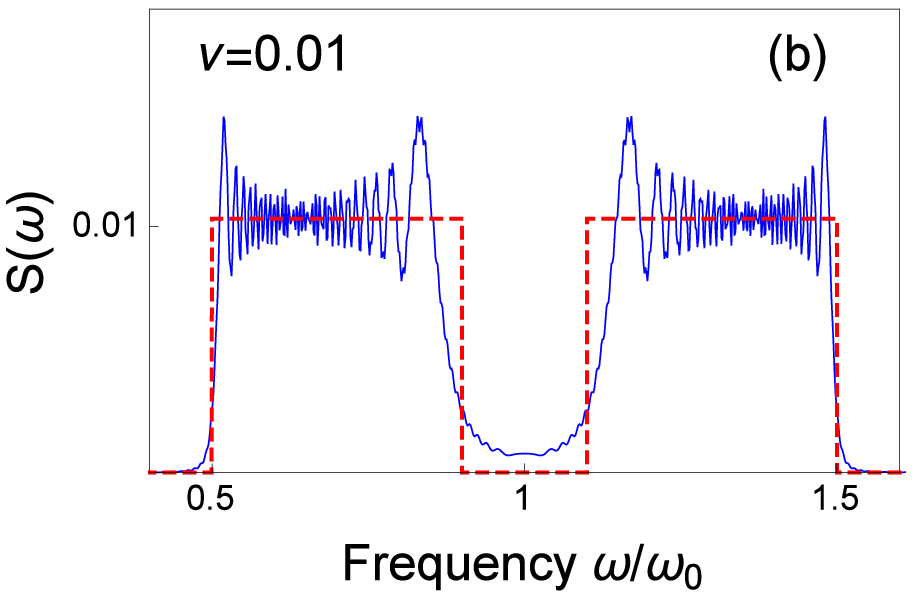}}\\
\subfloat{\includegraphics[width=.48\linewidth]{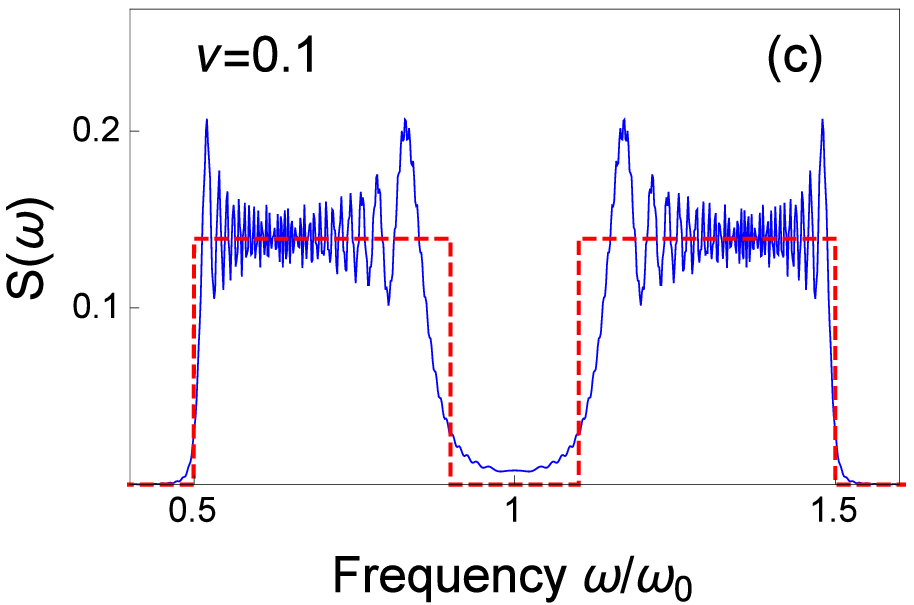}}\hspace{.02\linewidth}
\subfloat{\includegraphics[width=.48\linewidth]{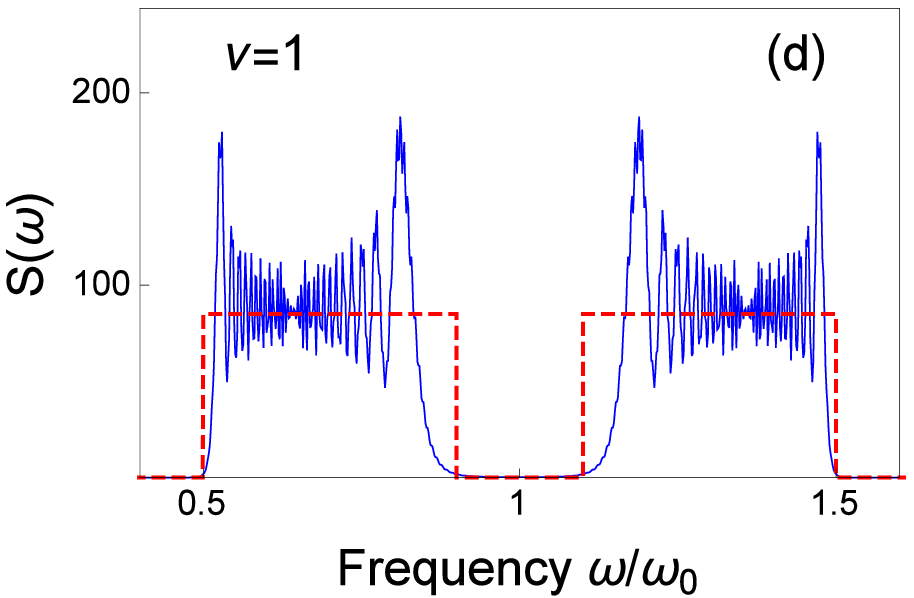}}
\caption{\label{OS} Optical spectra of the signal and the idler fields of PDC in an aperiodically poled crystal with a linear poling profile. Solid blue line, exact solution; dashed red line, first-order approximation. The parameter $\nu$ is proportional to the intensity of the pump.}
\end{figure}

In Fig.~\ref{OS} we show the PDC spectra, calculated with the help of the exact
solution, defined by Eqs.~(\ref{BogoliubovUV}), (\ref{U}), (\ref{A}), and (\ref{B}), and its approximation, Eq.~(\ref{Bogol2}), for various values of the parameter $\nu =4\left|\chi_{0}b_{p}\right|^2/(\pi^2\zeta)$.

We see in Fig.~\ref{OS} that for both small and comparable-with-unity values of parameter $\nu$ the approximation in Eq.~(\ref{Bogol2}) gives the true average value for the spectrum in the generation band, though it does not follow the rapid oscillations around this average value. Validity of the approximate formula at small values of $\nu$ is a consequence of unitarity of the field transformation in the amplification layer, Eq. (\ref{Bogol}). This is the key difference from the approach of Ref. \cite{Fejer08a}, where a purely classical consideration was undertaken, valid for a sufficiently strong pump, $\pi\nu\gg1$. It is demonstrated in the same reference that the rapid oscillations of the spectrum can be fairly well described by the second-order (WKB) approximation. For the purposes of design of aperiodically poled crystals these oscillations may be secondary and a general, ``averaged,'' shape of the spectrum provided by Eq.~(\ref{Bogol2}) may be sufficient.

The squeezing spectrum of the field at the crystal output $S_2(\Omega)$ for an unseeded PDC is defined by the relation $\langle X_2(\Omega,L)X_2^{\dagger}(\Omega',L)\rangle = S_2(\Omega)\langle X_2(\Omega,0)X_2^{\dagger}(\Omega',0)\rangle$; i.e., it shows the change of variance of the quadrature $X_2$. It is given by \cite{Kolobov99},
\begin{equation}\label{S2}
S_{2}(\Omega)=\left( \left| U(\Omega) \right|-\left| V(\Omega) \right| \right)^2.
\end{equation}

In Fig.~\ref{S2} we show the squeezing spectra, calculated on the basis of the exact
solution, defined by Eqs.~(\ref{BogoliubovUV}), (\ref{U}), (\ref{A}), and (\ref{B}), and its approximation, Eq.~(\ref{Bogol2}), for various values of the parameter $\nu$.

\begin{figure}[htbp]
\subfloat{\includegraphics[width=.48\linewidth]{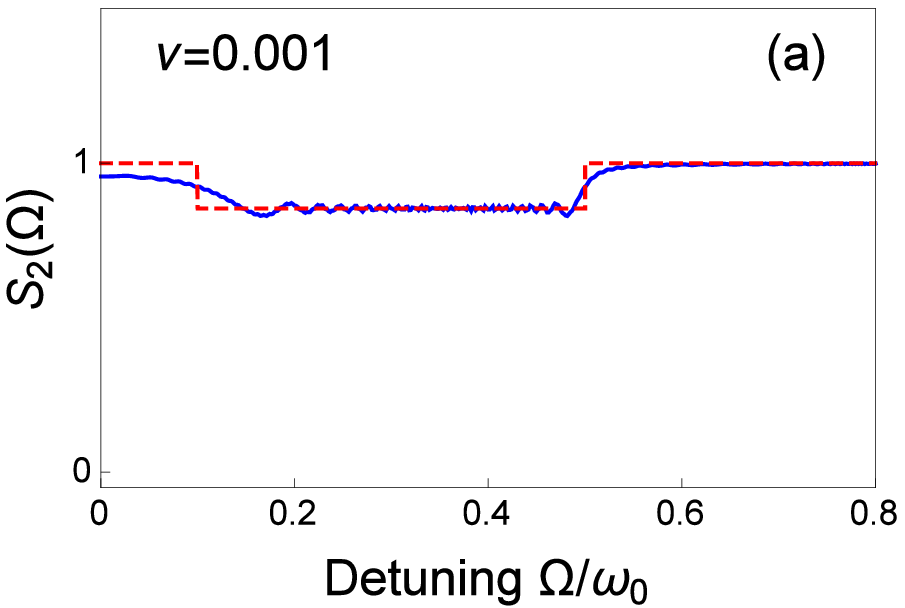}}\hspace{.02\linewidth}
\subfloat{\includegraphics[width=.48\linewidth]{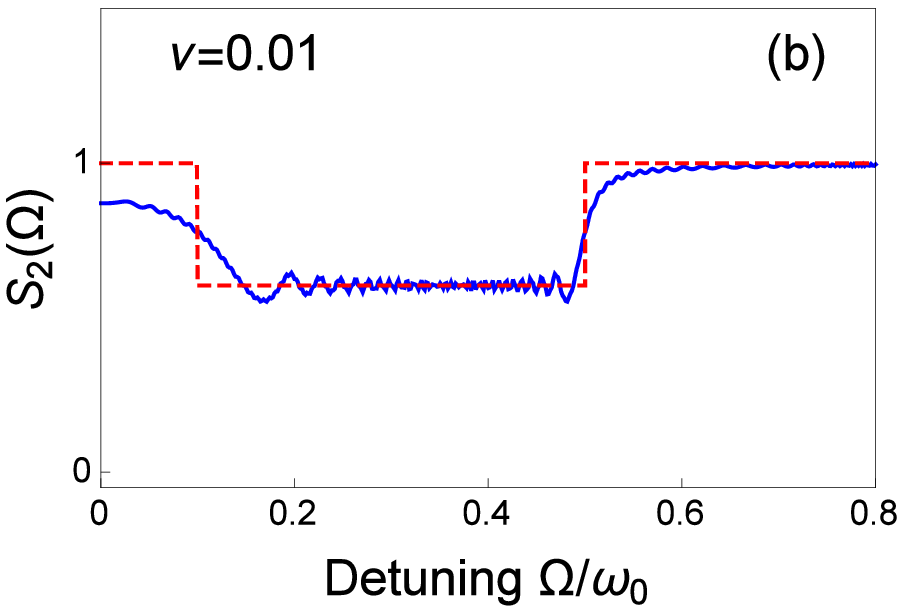}}\\
\subfloat{\includegraphics[width=.48\linewidth]{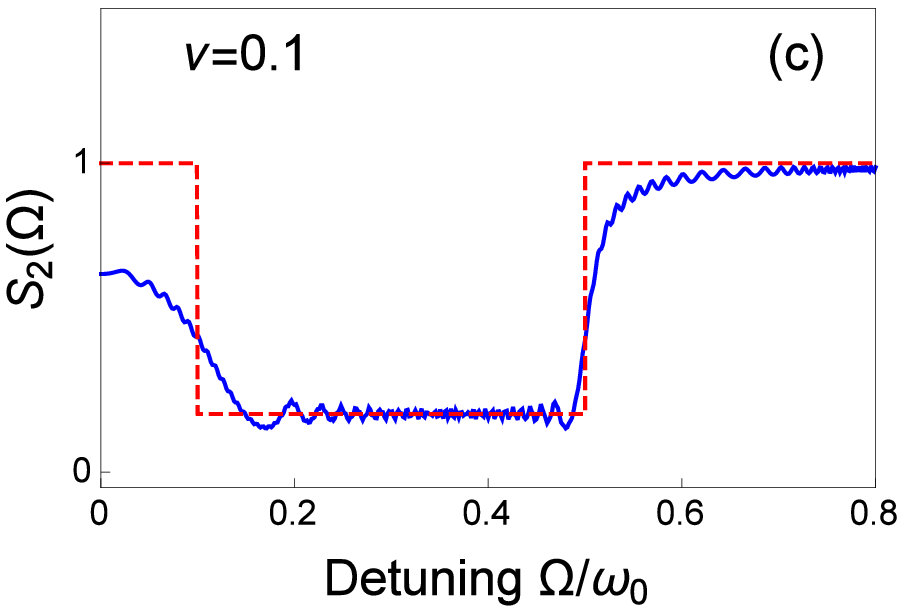}}\hspace{.02\linewidth}
\subfloat{\includegraphics[width=.48\linewidth]{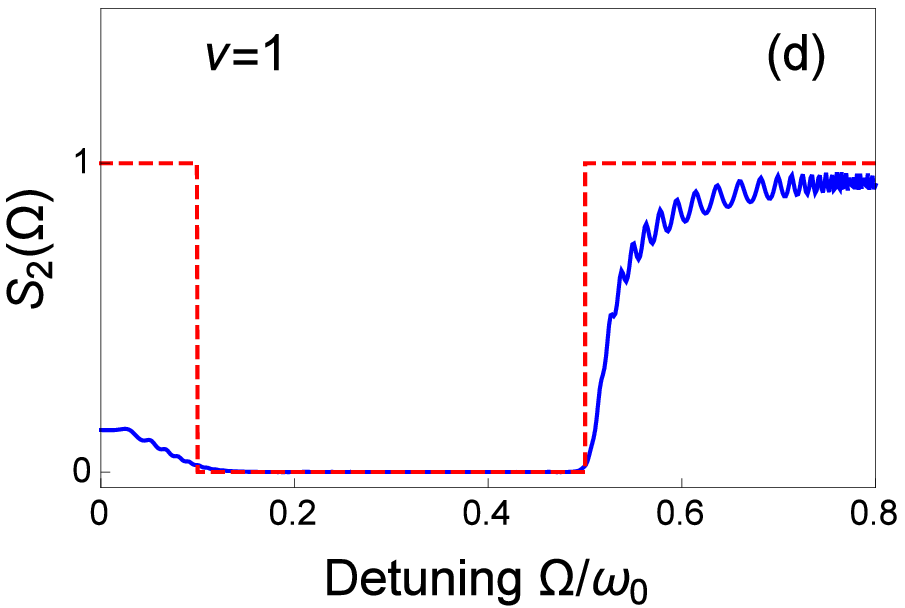}}
\caption{\label{S2} Squeezing spectra for a linear poling profile. Solid blue line, exact solution; dashed red line, first-order approximation.}
\end{figure}

We see in Fig.~\ref{S2} that for both small and comparable-with-unity values of parameter $\nu$ the approximation in Eq.~(\ref{Bogol2}) gives a value for the spectrum very close to the exact value in the generation band. Outside this band there is a significant difference between the two solutions: the exact solution shows some squeezing, though the approximate one predicts no squeezing at all.

The first-order approximate angle of squeezing can be easily obtained for a linear chirp by integrating Eq.~(\ref{varphi}) and substituting the result into Eq.~(\ref{psiL}), which  gives
\begin{equation}\label{psiLlin}
\psi_L^{(lin)}(\Omega)=\frac{1}{2}\left[(k(\Omega)+k(-\Omega))L - \frac{\left(\Delta(\Omega)-K_0\right)^2}{2\zeta}+ \bar\varphi_A \right],
\end{equation}
where $\bar\varphi_A=\varphi_A-2k_0L$. A similar expression (with a factor of $-2$) was obtained in Ref.~\cite{Harris07} in the low-gain regime for the phase of the compensating optical element, necessary to provide simultaneous arrival of the signal and idler photons at the distant detector or summing crystal. Thus the angle of squeezing is related to this phase as $\psi_L(\Omega)=-\frac12\arg\{H(\Omega)\}+\frac12\bar\varphi_A$, where $H(\Omega)$ is the transfer function of the compensating element. The additional term $2k_0L$ in the definition of $\bar\varphi_A$ reflects the difference between the phase of a sideband operator and that of the full field operator. The phase $\varphi_1$, which is part of $\varphi_A$, is very small in the low-gain regime and can be disregarded. In the high-gain regime it is not small, but for a linear chirp it is constant and does not affect the simultaneity of photon arrivals. Thus, even in the high-gain regime we can understand the dispersion of the squeezing angle as caused by a relative delay of the photons at the conjugated frequencies.

In a similar way we obtain the second characteristic angle,
\begin{equation}\label{psi0lin}
\psi_0^{(lin)}(\Omega)=-\frac{\left(\Delta(\Omega)-K_0\right)^2}{4\zeta}+\varphi_A.
\end{equation}
We see that in the first-order approximation both angles at all frequencies are independent of the pump power up to an additive constant, which is quite a general result.

\begin{figure}[htbp]
\includegraphics[width=\linewidth]{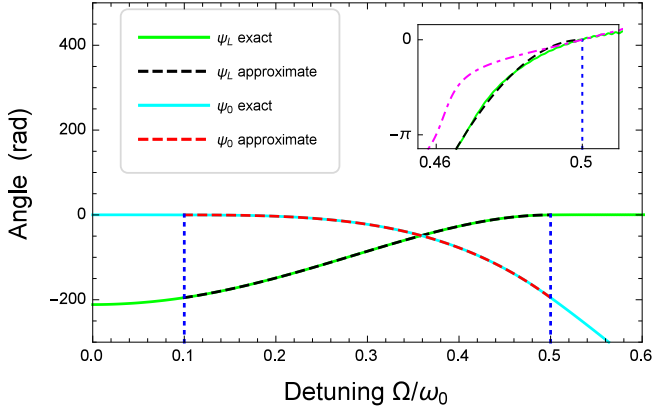}
\caption{\label{AnglesExact} Angles $\psi_L(\Omega)$ and $\psi_0(\Omega)$ for a linear poling profile at the pump power corresponding to medium squeezing, $\nu=0.01$. Solid green line (increasing), exact solution for $\psi_L(\Omega)$; dashed black line (increasing), its first-order approximation. Solid cyan line (decreasing), exact solution for $\psi_0(\Omega)$; dashed red line (decreasing), its first-order approximation. $\varphi_A$ is chosen so that $\psi_L(0.5\omega_0)=0$. Vertical dotted lines mark the borders of the amplification band. Inset: Dash-dotted magenta line shows the exact values of $\psi_L(\Omega)$ at $\nu=1$, which are already significantly different from the first-order approximation.}
\end{figure}

We show in Fig.~\ref{AnglesExact} the exact and the approximate values for both angles, calculated at $\nu=0.01$. We see that the difference between the two solutions is much smaller than $\pi/2$ everywhere. With the growing $\nu$ this difference increases, and a numerical study shows that up to $\nu=1$ it remains less than or comparable to $\pi/2$.

We can conclude that the approximate formula, Eq.~(\ref{Bogol2}), looks very promising for designing aperiodically poled crystals in the case of a linear chirp profile. It provides a good qualitative description of the squeezing spectrum and almost exact values of the characteristic angles for a sufficiently low pump power, $\nu$ well below 1.

\subsection{Nonlinearly chirped crystal: comparison of approximate and numerical solutions \label{spectranl}}

In this section we analyze a nonlinear profile of aperiodic poling. Several specific shapes of nonlinear profiles have been studied to date: a $z^n$ profile \cite{Baker03}, and sinusoidal and tapered profiles \cite{Fejer08a}. For demonstrating the efficiency of the results of Sec.~\ref{approx} we need to consider a rather slowly varying profile, which is selected from the following physical considerations. In the previous section we have seen that a linear profile produces an almost flat spectrum of the down-converted light, but the angle of squeezing, $\psi_L(\Omega)$, is a complicated function of frequency detuning, including non-negligible third- and fourth-order components [see Eq.~(\ref{psiLlin})]. Any observation of the ultrabroadband character of squeezing requires compensation of this angle in a wide range of frequencies. For a bandwidth of the order of 10 THz the quadratic term is dominant and compensation can be performed by a passive optical element (optical fibre \cite{Brida09}, a glass block \cite{Sensarn10}, or a pair of prisms \cite{Tanaka12}), but for a 100-THz-wide PDC spectrum an active compensation is required \cite{Peer05,Lukens13}, which is the state of art of modern quantum optics. When looking for a nonlinear spatial frequency profile we could demand that it is a monotonous function $K(z)$ such, that the corresponding angle of squeezing $\psi(\Omega)$ is a second-order polynomial, which can be compensated by passive optical elements.

The relative delay of the signal with respect to the idler in the first-order approximation is given by Eq.~(\ref{psidiff}). Our aim is to obtain $\tau(\Omega)=a\Omega+b$, where $a$ and $b$ are some real parameters. From Eqs.~(\ref{zpm}), (\ref{psidiff}), and (\ref{ang}) we obtain the following equation, which should be satisfied by the profile function:
\begin{equation}\label{eq}
 K\left(L-\frac{a\Omega+b}{\Delta'(\Omega)}\right)=\Delta(\Omega).
\end{equation}
This equation can be easily solved in the approximation of quadratic dispersion, Eq.~(\ref{Dkq}), where the inverse group velocity difference has a simple form: $\Delta'(\Omega)=-2\alpha\Omega/\omega_0^2$.

For a quadratic phase (linear delay) we need $z_{pm}(\Omega)=L+d+db/(a\Omega)$, where $d=a\omega_0^2/(2\alpha)$. The inverse of this function is
\begin{equation}\label{Omega}
\Omega_{pm}(z)=-\frac{db/a}{L+d-z},
\end{equation}
and has a meaning of the frequency, for which perfect phase matching is reached at the point $z$. The sought profile is found in the form
\begin{equation}\label{hyperb}
K(z)=\Delta(\Omega_{pm}(z))=-\frac\alpha{\omega_0^2}\left(\frac{db/a}{L+d-z}\right)^2+\beta.
\end{equation}

Let us determine the possible values of the coefficients $a$ and $b$. From Eq.~(\ref{Omega}) we obtain the phase-matched frequencies at the edges of the crystal,

\begin{equation}\label{lim}
\Omega_{pm}(0)=-\frac ba\frac{d}{L+d},\quad \Omega_{pm}(L)=-\frac ba.
\end{equation}
Since both frequencies should be positive, we have two possibilities:
\begin{itemize}
\item $a>0$, $b<0$, $d>0$, and $K(z)$ is decreasing, and
\item $a<0$, $b>0$, $d<-L$, and $K(z)$ is increasing.
\end{itemize}

In this section we limit the lower frequency of the signal amplification band to $0.25\omega_0$, because otherwise the profile does not satisfy the requirement of slow variation. In the first of the cases listed above, substituting $\Omega_{pm}(0)=0.25\omega_0$, $\Omega_{pm}(L)=0.5\omega_0$, we obtain $d=L$ and
\begin{equation}\label{hyperb2}
K(z)=-\frac{\alpha}{4\left(2-z/L\right)^2}+\beta.
\end{equation}
In the second case, substituting $\Omega_{pm}(0)=0.5\omega_0$, $\Omega_{pm}(L)=0.25\omega_0$, we obtain $d=-2L$ and
\begin{equation}\label{hyperb3}
K(z)=-\frac{\alpha}{4\left(1+z/L\right)^2}+\beta.
\end{equation}

\begin{figure}[htbp]
\subfloat{\includegraphics[width=.48\linewidth]{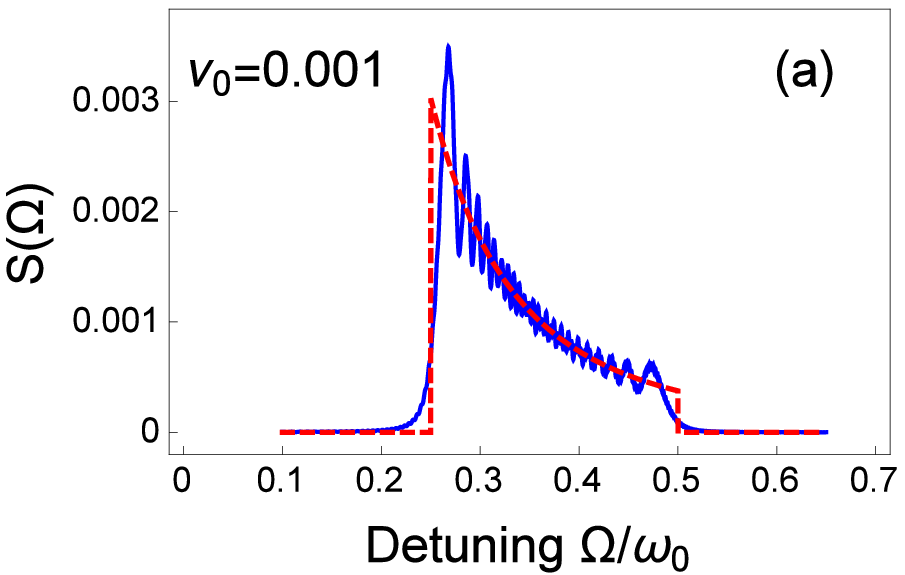}}\hspace{.02\linewidth}
\subfloat{\includegraphics[width=.48\linewidth]{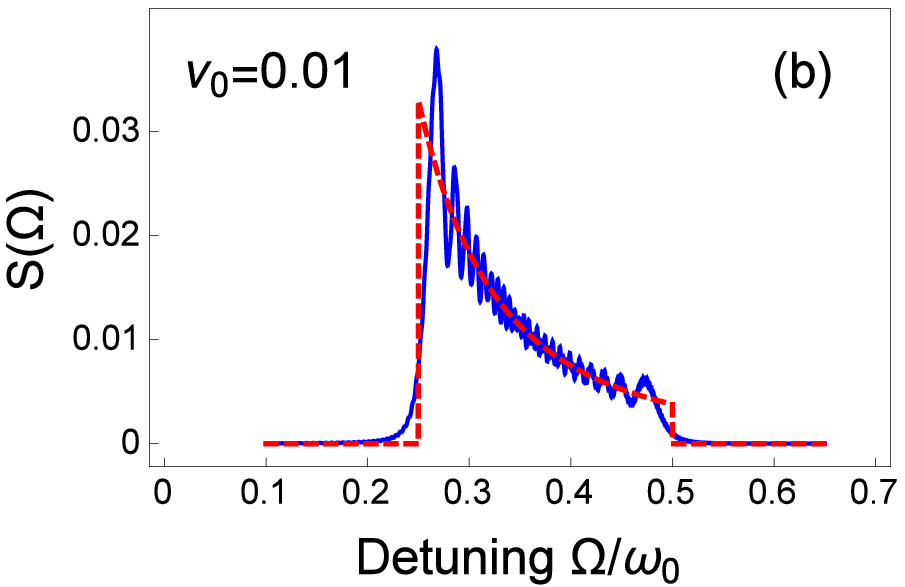}}\\
\subfloat{\includegraphics[width=.48\linewidth]{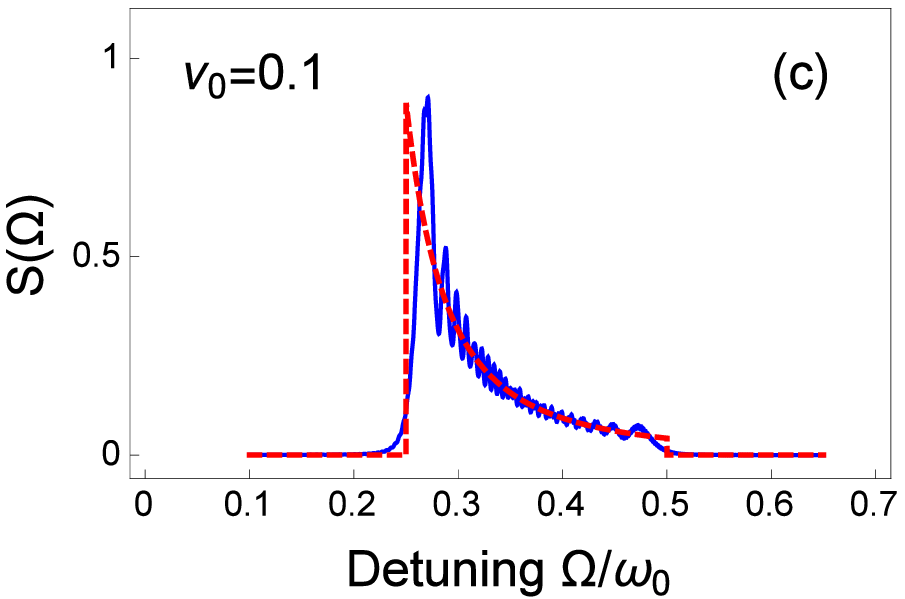}}\hspace{.02\linewidth}
\subfloat{\includegraphics[width=.48\linewidth]{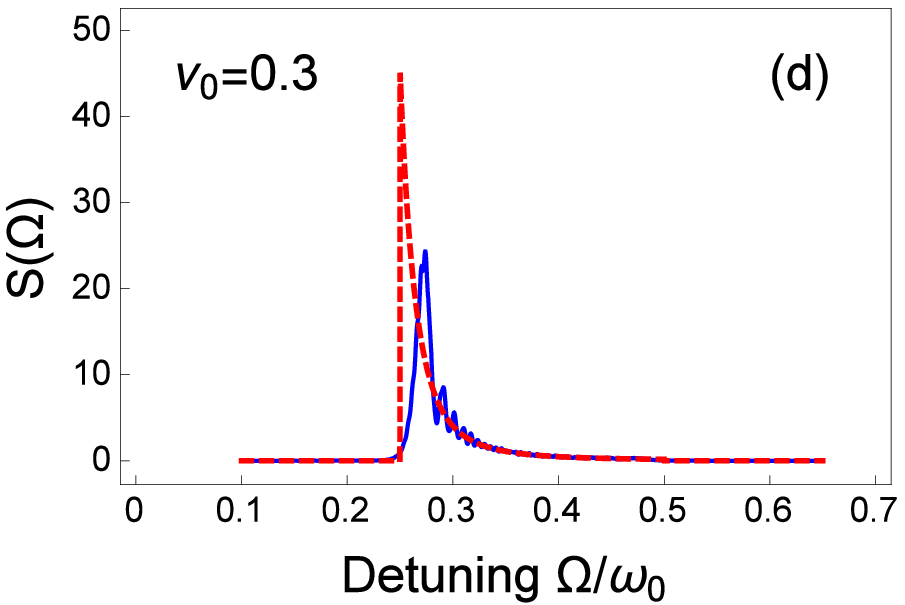}}
\caption{\label{OSnl} Optical spectra for the case of a quadratic-hyperbolic profile. Solid blue line, numerical solution; dashed red line, first-order approximation. The parameter $\nu_0$ is proportional to the intensity of the pump.}
\end{figure}

\begin{figure}[htbp]
\subfloat{\includegraphics[width=.48\linewidth]{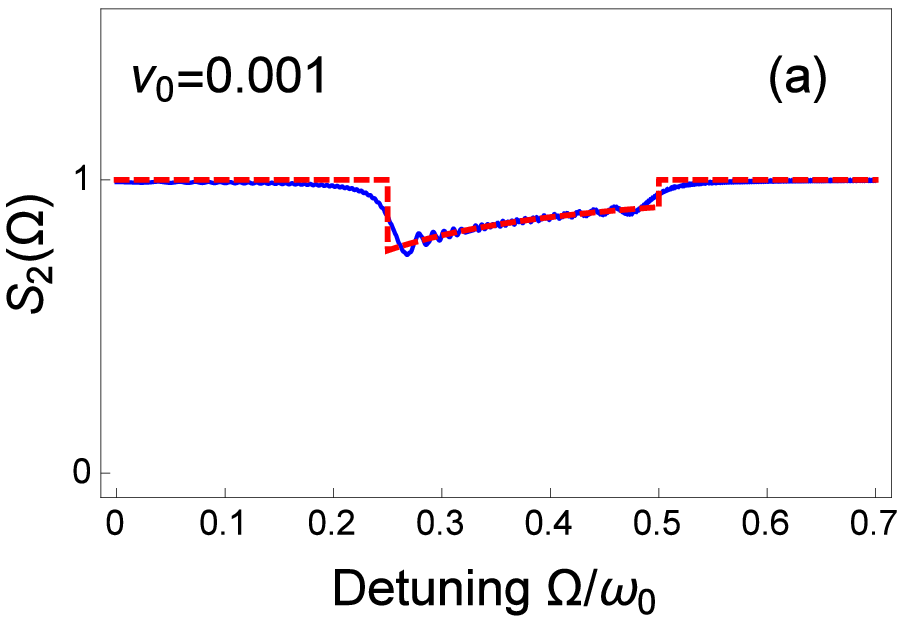}}\hspace{.02\linewidth}
\subfloat{\includegraphics[width=.48\linewidth]{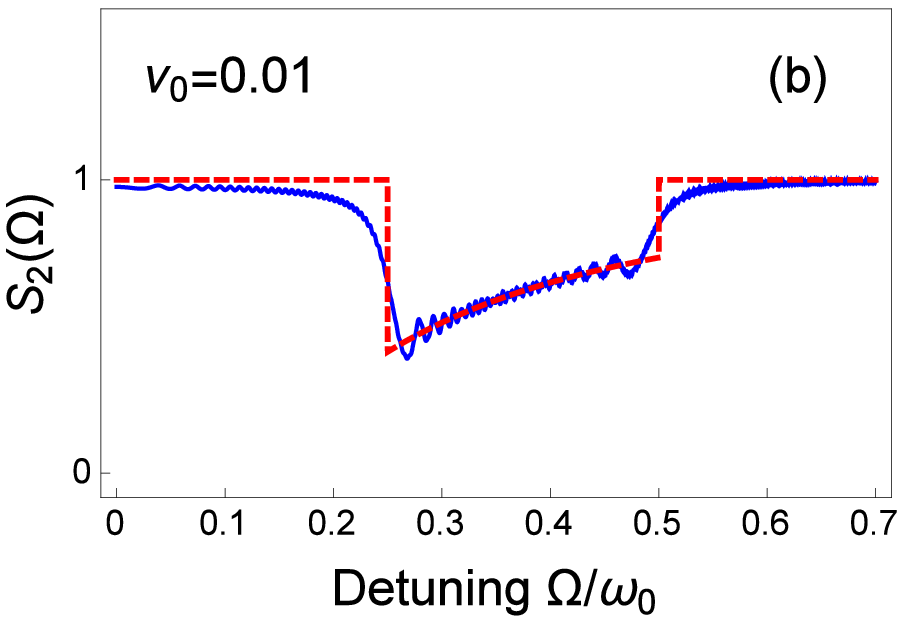}}\\
\subfloat{\includegraphics[width=.48\linewidth]{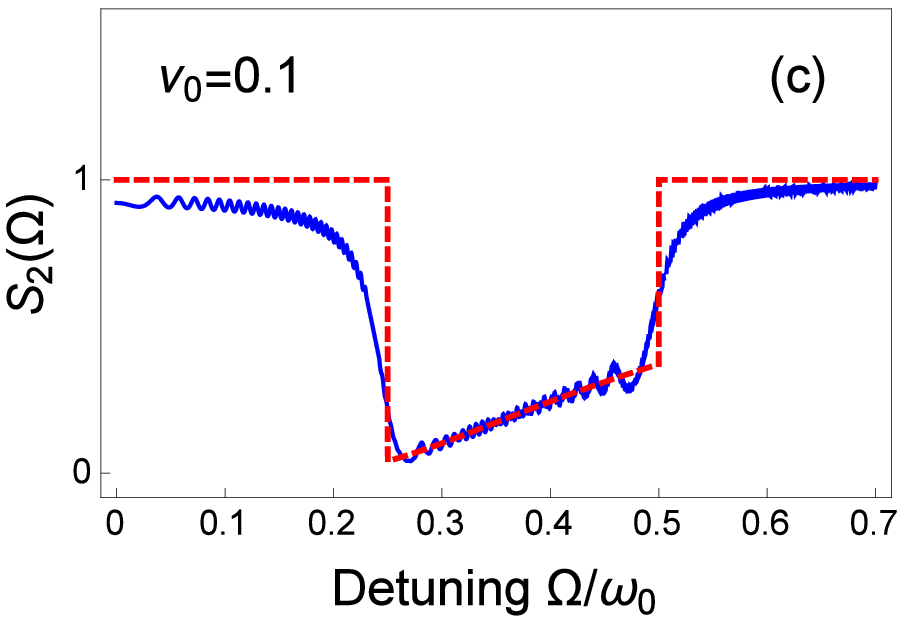}}\hspace{.02\linewidth}
\subfloat{\includegraphics[width=.48\linewidth]{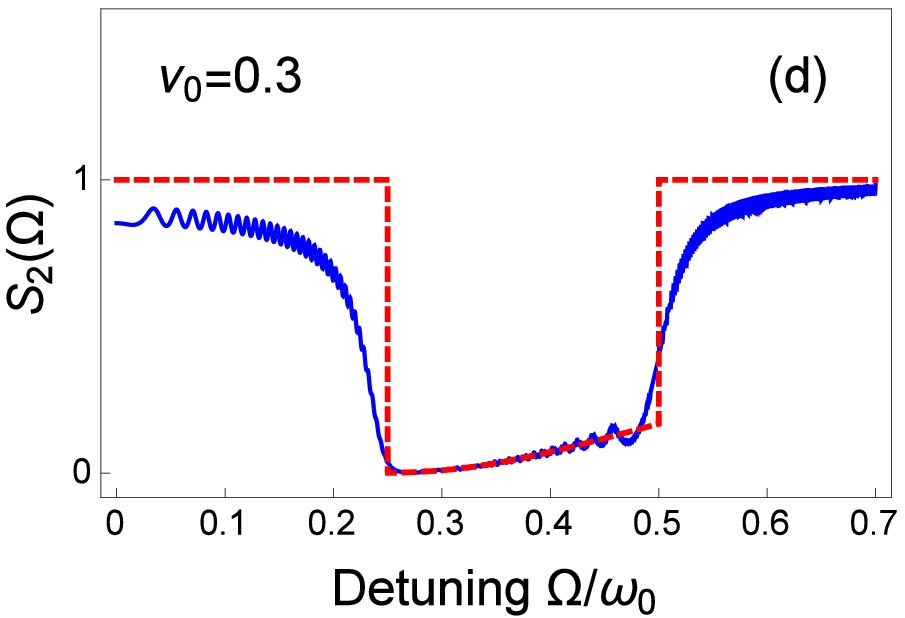}}
\caption{\label{S2nl} Spectra of squeezing for the case of quadratic-hyperbolic profile. Solid blue line, numerical solution, dashed red line, first-order approximation. The parameter $\nu_0$ is proportional to the intensity of the pump.}
\end{figure}

A decreasing profile is more interesting from the practical point of view, because it generates a negatively chirped field ($a>0$), where lower signal frequencies are more delayed than the higher ones, which requires a compensating medium with normal dispersion, e.g., an optical fiber \cite{Brida09} or a glass block \cite{Sensarn10}. In the rest of this section we compare the first-order approximate and numerical solutions for the case of decreasing quadratic-hyperbolic poling profile, given by Eq.~(\ref{hyperb2}). Substituting Eq.~(\ref{hyperb2}) and Eq.~(\ref{Dkq}) into Eq.~(\ref{waveeq6}) and solving numerically this second-order differential equation, we calculate optical spectra and spectra of squeezing. These spectra are presented in Figs.~\ref{OSnl} and \ref{S2nl}, where they are compared with the first-order values, predicted by Eq.~(\ref{Bogol2}). For better comparison with Figs.~\ref{OS} and \ref{S2}, we introduce normalized pump intensity,
\begin{equation}
\nu_0=\frac{|\gamma|^2L}{|K(0)-K(L)|},
\end{equation}
which has a physical meaning of the Rosenbluth parameter for the linear profile, providing a quasi-phase-matching in the same frequency band for the given crystal length.

\begin{figure}[htbp]
\centerline{\includegraphics[width=\linewidth]{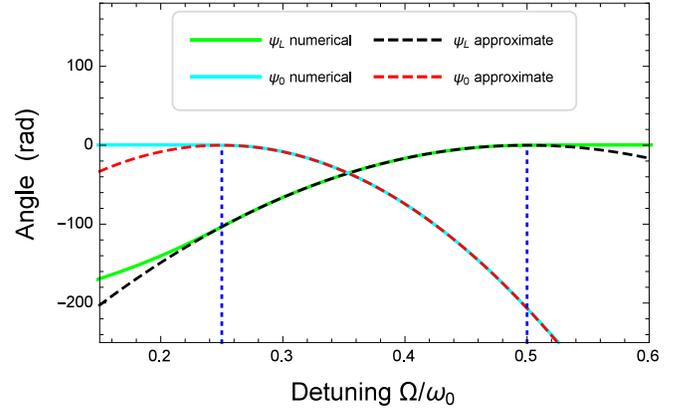}}
\caption{\label{AnglesNum} Angles $\psi_L(\Omega)$ and $\psi_0(\Omega)$ for a quadratic-hyperbolic poling profile at the pump power corresponding to $\nu_0=0.01$. Solid green line (increasing), numerical solution for $\psi_L(\Omega)$; dashed black line (increasing), its first-order approximation. Solid cyan line (decreasing), numerical solution for $\psi_0(\Omega)$; dashed red line (decreasing), its first-order approximation. All angles are shifted so that $\psi_L(0.5\omega_0)=\psi_0(0.25\omega_0)=0$. Vertical dotted lines mark the borders of the amplification band. }
\end{figure}

From the same numerical solution we can calculate the angle of squeezing. Its first-order value is given by integrating Eq.~(\ref{varphi}) with the profile defined by Eq.~(\ref{hyperb2}) and substituting the result into Eq.~(\ref{psiL}),
\begin{equation}\label{psiLhyp}
\psi_L^{(qh)}(\Omega)=-\frac{\alpha L}{2}\left(\frac{\Omega-0.5\omega_0}{\omega_0}\right)^2+\psi_{\mathrm{c}},
\end{equation}
where $\psi_{\mathrm{c}}=(\varphi_A+\alpha L/8-\beta L)/2$ is a constant. In a similar way we obtain the input angle
\begin{equation}\label{psi0hyp}
\psi_0^{(qh)}(\Omega)=-\alpha L\left(\frac{\Omega-0.25\omega_0}{\omega_0}\right)^2+\frac{\varphi_A}2,
\end{equation}
where the superscript $qh$ denotes the quadratic-hyperbolic profile. The approximate angles are plotted in Fig.~\ref{AnglesNum} together with their numerical solutions. We see that the agreement of both solutions is very good. As in the previous section, the numerical study shows that the two solutions start to differ significantly when $\nu_0$ approaches unity.

Finally, let us evaluate two parameters, characterizing the slow variation of the profile, given by Eq.~(\ref{hyperb2}). We find easily that $\max|\Lambda'(z)|= |\Lambda'(L)|=0.001$, so the poling period is changing sufficiently slowly for applying Eq.~(\ref{meander2}). We also find that $\max(\epsilon')=0.18\sqrt{\nu_0}$, and, therefore, the linearization of the poling profile is justified for values $\nu_0<0.31$, where $\epsilon'<0.1$. We see in Fig.~\ref{OSnl}(d) that at $\nu_0=0.3$ the optical spectrum starts to deflect from the prediction of the first-order approximation not only in rapid oscillations but also in the average value. With growing $\nu_0$ this deflection becomes greater, meaning that the linearized solution is not valid anymore.

To ascertain that the good correspondence of the approximate solution to the exact analytic and the numerical ones is not a particular property of the chosen crystal settings, we applied the analysis of the current section to four other crystal designs. First, we considered MgO:LiNbO$_3$ crystals of different lengths, quasi-phase-matched for the same bandwidth: a 2-mm-long crystal with the chirp rate $\zeta=87$ rad/mm$^2$, and a 20-mm-long crystal with the chirp rate $\zeta=8.7$ rad/mm$^2$. In addition, we considered a 20-mm-long crystal of undoped LiNbO$_3$ pumped at 420 nm and quasi-phase-matched from 464 to 750 nm, as in Ref.~\cite{Harris07}, and a 22-mm-long crystal of stoichiometric LiTaO$_3$ pumped at 532 nm and quasi-phase-matched from 680 to 800 nm, as in Ref.~\cite{Fejer05}. In all these cases we obtained the correspondence of the solutions similar to that of Figs.~5--10.

\section{Conclusions \label{conclusion}}

We have considered the process of ultrabroadband collinear PDC in an aperiodically poled crystal, designed to produce QPM in a wide range of wavelengths (hundreds on nanometers). In the case of the high-gain regime with an undepleted pump such a process generates an ultrabroadband squeezed-light wave at the output of the crystal. The components of such a light wave at the frequencies symmetric with respect to the central frequency $\omega_0$ are highly quantum correlated, and their correlation time may be made as small as one optical period. This ultrabroadband squeezing can be observed, for example,  in second-harmonic generation as described in Ref.~\cite{Horoshko13}, after the compensation of the angle of squeezing at all frequencies. For a sufficiently broadband squeezed light the correlation time can be as short as a single optical period.

We can estimate the number of squeezed modes in the considered ultrabroadband source of squeezed light. In our model of the monochromatic pump the number of squeezed modes is formally infinite. When the spectral width $\delta \omega$ of the pump is taken into account, the number of such modes in the low-gain regime is given roughly by the ratio $\Delta \omega/\delta \omega$, where $\Delta \omega$ is the amplification bandwidth \cite{Horoshko12}. In the high-gain regime the modes are expected to be approximately the same, but each pair of modes will be characterized by a high degree of squeezing. Thus, we can estimate the number of entangled modes for a nanosecond pump pulse as $200 \mathrm{THz}/1\mathrm{GHz}=2\times10^5$ modes, where the amplification bandwidth of 200 THz corresponds to the example analyzed in Sec.~\ref{spectra}. Such a highly multimode field can be used in various applications of quantum information, from metrology to cluster state quantum computation.

Let us summarize the results obtained in this article. First, we have analyzed in detail the exact solution of the differential equation for PDC with an undepleted quasi-monochromatic pump in an aperiodically poled nonlinear crystal with a linear poling profile. The solution is expressed through parabolic cylinder special functions. We have analyzed the properties of this solution and proven the conservation of the commutation relations for the field operators. Second, we have obtained a unitary approximate solution, a ``quantum Rosenbluth formula,'' in the first-order approximation and have demonstrated that it is in good agreement with the exact solution within the amplification band for various values of the pump power. We have shown that, taking into consideration the quantum conditions, one arrives at a solution, applicable in the high-gain regime of PDC with the gain, corresponding to practical values of squeezing from 3 to 12 dB. We have also shown a good correspondence of the approximate solution to the numerical one for the case of a nonlinear (quadratic-hyperbolic) profile. Third, we have shown that a quadratic-hyperbolic profile of aperiodic poling results in a negatively chirped output field, compressible by a passive dispersive element with normal quadratic dispersion. These results will help to design aperiodically poled crystals for generation of squeezed light with monocycle squeezing, which is important for applications requiring ultra-short correlations in the temporal domain or an ultra-high number of entangled modes in the spectral domain.

\acknowledgments

The authors are grateful to Maria Chekhova and Chris Phillips for fruitful discussions. This work was supported in part by the European Union's Horizon 2020 research and innovation programme under grant agreement No 665148 (QCUMbER) and in part by the Belarusian Republican Foundation for Fundamental Research.

\end{document}